\documentclass[authoryear,3p,10pt]{elsarticle}
\usepackage{graphicx}
\usepackage{tabulary}
\usepackage{epsfig}
\usepackage{amssymb}
\usepackage{multirow}
\usepackage{appendix}

\newcommand{\ME}{\,M_{\oplus}}
\newcommand{\Ms}{M_{\star}}
\newcommand{\beq}[1]{\begin{equation}\label{#1}}
\newcommand{\eeq}{\end{equation}}
\newcommand{\CTE}{CTE\ }
\newcommand{\MTE}{MTE\ }
\newcommand{\Rol}{Ro^*_l}
\newcommand{\mc}[1]{\multicolumn{2}{c}{#1}}
\newcommand{\mr}[1]{\multirow{2}{*}{#1}}
\renewcommand{\bar}{\overline}
\journal{Icarus}
\begin{document}

\begin{frontmatter}

\title{The role of rotation in the evolution of dynamo-generated
  magnetic fields in Super Earths}

\author{Jorge I. Zuluaga}
\ead{jzuluaga@fisica.udea.edu.co}
\author{Pablo A. Cuartas}
\ead{p.cuartas@fisica.udea.edu.co}

\address{Instituto de F\'\i sica - FCEN, Universidad de
  Antioquia, Calle 67 No. 53-108, Medell\'\i n, Colombia}

\begin{abstract}
Planetary magnetic fields could impact the evolution of planetary
atmospheres and have a role in the determination of the required
conditions for the emergence and evolution of life (planetary
habitability).  We study here the role of rotation in the evolution of
dynamo-generated magnetic fields in massive Earth-like planets, Super
Earths (1-10 $\ME$).  Using the most recent thermal evolution models
of Super Earths \citep{Gaidos10, Tachinami10} and updated scaling laws
for convection-driven dynamos, we predict the evolution of the local
Rossby number. This quantity is one of the proxies for core magnetic
field regime, i.e. non-reversing dipolar, reversing dipolar and
multipolar.  We study the dependence of the local Rossby number and
hence the core magnetic field regime on planetary mass and rotation
rate.  Previous works have focused only on the evolution of core
magnetic fields assuming rapidly rotating planets, i.e. planets in the
dipolar regime.  In this work we go further, including the effects of
rotation in the evolution of planetary magnetic field regime and
obtaining global constraints to the existence of intense protective
magnetic fields in rapidly and slowly rotating Super Earths.  We find
that the emergence and continued existence of a protective planetary
magnetic field is not only a function of planetary mass but also
depend on rotation rate.  Low-mass Super Earths ($M\lesssim 2\ME$)
develop intense surface magnetic fields but their lifetimes will be
limited to 2-4 Gyrs for rotational periods larger than 1-4 days.  On
the other hand and also in the case of slowly rotating planets, more
massive Super Earths ($M\gtrsim 2\ME$) have weak magnetic fields but
their dipoles will last longer.  Finally we analyze tidally locked
Super Earths inside and outside the habitable zone of GKM stars.
Using the results obtained here we develop a classification of Super
Earths based on the rotation rate and according to the evolving
properties of dynamo-generated planetary magnetic fields.
\end{abstract}


\end{frontmatter}


\section{Introduction}
\label{sec:introduction}

The number of known exoplanets in the mass range between 1 and 10 $\ME$
is growing (hereafter these objects will be called ``Super Earths'' or
SEs following the classification by
\cite{Valencia06,Valencia07a}). At the time of writing, there are
almost 46 confirmed planets in this mass range\footnote{For updates,
  please refer to {\tt http://exoplanet.eu}}\citep{Rivera05,
  Beaulieu06, Udry07, Mayor08, Ribas08, Queloz09, Bonfils11,
  Lissauer11} and more than a few hundred SEs candidates are awaiting
further analysis and confirmation \citep{Borucki11}.  These
discoveries have increased the interest to model and understand the
geophysical properties of this type of planets \citep{Valencia06,
  Valencia07a, Valencia07b,Valencia09,Seager07,Kaltenegger10a,Korenaga10}.  The
habitability of SEs, in particular those similar in composition and
structure to the Earth, is an interesting topic in the field and
several theoretical works have paid special attention to this
particular aspect of SEs properties \citep{Griebmeier05, Griebmeier09,
  Griebmeier10, Selsis07a, VanThienen07, VonBloh07, Lammer10}.

Models of the interior structure of SEs have been extensively
developed over the last 5 years \citep{Valencia06, Valencia07a,
  Valencia07b, Fortney07, Seager07, Selsis07a, Sotin07, Adams08,
  Baraffe08, Grasset09}.  Although there are still open issues to be
addressed, these models are giving us an understanding of global
properties such as the mass-radius relationship and its dependence
with planetary composition, as well as different geophysical phenomena
such as mantle convection, degassing and plate tectonics
\citep{Olson07,Papuc08,Valencia07c,Valencia09,Korenaga10}.  Recently
several authors have studied in detail the thermal evolution and
magnetic field properties of this type of planets
\citep{Gaidos10,Tachinami10, Driscoll11}.

Planetary magnetic fields would likely play a role in planetary
habitability \citep{VonBloh07, Griebmeier05, Griebmeier09,
  Griebmeier10, VanThienen07, Lammer10}.  Understanding the conditions
for the emergence and long term evolution of a protective planetary
magnetic field (hereafter PMF) is crucial to evaluate the complex
conditions for habitability of SEs.  The same conditions could also be
applied to evaluate the habitability of exomoons around extrasolar
giant planets \citep{Kaltenegger10b}.

The current understanding of PMF emergence and evolution in SEs arises
from thermal evolution models for the Earth \citep{Stevenson03,
  Labrosse03, Labrosse07a, Labrosse07b, Nimmo09a, Aubert09, Breuer10}
and scaling laws for convection-driven dynamos obtained from extensive
numerical simulations \citep{Christensen06, Olson06, Aubert09,
  Christensen09, Christensen10}.  Two recent works studied the problem
of PMF evolution in SEs by developing detailed models of planetary
thermal evolution \citep{Gaidos10,Tachinami10}. Both works have paid
special attention to different but complementary aspects of the
problem.  \citet{Gaidos10} uses a model of the structure of the
planetary core and its thermal evolution (hereafter the {\it Core
  Thermal Evolution} or {\it \CTE model}).  On the other hand
\citet{Tachinami10} uses the Mixing Length Theory adapted to planetary
conditions to model mantle convection with a detailed treatment of its
rheological properties (hereafter the {\it Mantle based Thermal
  Evolution model} or {\it \MTE model}).  Although thermal evolution
models for the Earth, other terrestrial planets and even SEs have been
developed in the past \citep{Stevenson03, Labrosse03, Labrosse07a,
  Papuc08, Nimmo09a, Breuer10}, the \CTE and \MTE models give the
first detailed description aimed at studying dynamo-generated magnetic
fields of extrasolar terrestrial planets.

We use the results of the \CTE and \MTE models to study the role of
rotation in the evolution of PMF in SEs.  We focus on the evolution of
the regime of the core magnetic field (CMF) which can be broadly
classified as non-reversing dipolar, reversing dipolar and multipolar.
For this purpose we compute the {\it local Rossby number}, one of the
proxies for CMF regime, as a function of the rotation period and
planetary mass.  Using this property we predict the long term
evolution of the surface PMF in rapidly and slowly rotating planets.

In section \ref{sec:thermal-models} we summarize the most important
results of the \CTE and \MTE models.  Section \ref{sec:dynamo-props}
presents the scaling laws for convection-driven dynamos used to
predict the properties of the CMF.  In section \ref{sec:rotation-PMF}
we present a general procedure to compute the CMF intensity in the
dipolar and multipolar regimes including an implicit dependence on
rotation rate.  In section \ref{sec:results} we present the results of
applying the procedure devised here to predict the maximum dipolar
component of the field in SEs with different periods of rotation and
thermal histories as predicted by the \CTE and \MTE models.  Section
\ref{sec:discussion} is devoted to discuss the limitations of our
procedure and the implications of our results.  A summary, concluding
remarks and future prospects are presented in section
\ref{sec:conclusions}.  For reference a list of the symbols and the
physical quantities used in this work are presented in Table
\ref{tab:symbols}.

\section{Thermal evolution models of SEs}
\label{sec:thermal-models}

\subsection{Planetary thermal evolution}
\label{subsec:dynamo-thermal}

The evolution and long-term survival of a dynamo-generated
PMF\footnote{It is important to stress that we are considering in this
  work magnetic fields generated by dynamo action in a liquid metallic
  core.  Other fluid shells (liquid, ice or gaseous mantles) may
  sustain other type of dynamos out of the scope of this work.}
strongly depends on the thermal history of the planet.  The energy
sources and the amount of energy available for dynamo action change in
time as the planet cools and, in some cases, its core solidifies.

Before we present the predicted properties of planetary dynamos and
the role of rotation in the evolution of those properties, we present
the most recent advances in the study of thermal evolution of SEs.
Thermal evolution models are the starting point of any dynamo
evolutionary model. The onset of dynamo action in the core requires
values of the magnetic Reynolds number beyond a given threshold
$Re_m>40$ \citep{Olson06}.  The strength of the dynamo and hence the
intensity of the generated magnetic field are essentially set by the
available convective power $Q_{conv}$ \citep{Olson06}.  An important
role is also played by the size of the convective region that in core
dynamos is given by the difference between the radius of the outer
core and the radius of the solid inner core.  In the thermal evolution
models studied here \citep{Gaidos10,Tachinami10} $Q_{conv}$ is
estimated in two different ways.  In the \CTE model this quantity is
obtained by solving first the thermal equilibrium equations for the
core and mantle and then computing the ohmic dissipation from the
balance between entropy sources (radioactive decay, secular cooling,
sensible heat, latent heat and other sources of buoyancy) and sinks
(heat conduction, ohmic dissipation and other sources of dissipation)
\citep{Lister03,Labrosse03,Labrosse07a,Nimmo09a,Gaidos10}.  In the
\MTE model, $Q_{conv}$ is estimated solving first the thermal
transport equations in the mantle and the core and computing the
difference between the heat that comes out of the core-mantle boundary
(CMB) and the amount of energy transported by conduction through it
\citep{Tachinami10}.  Important differences between the \CTE and \MTE
models arise from the way $Q_{conv}$ is estimated (see section
\ref{subsec:MTEmodel}).  In the following sections we discuss in
detail some of the specific features of the \CTE and \MTE models.

\subsection{The \CTE model}
\label{subsec:CTEmodel}

\citet{Gaidos10} (\CTE model) solved the entropy equilibrium equations
in the core assuming adequate initial and boundary conditions and a
parametrized description of the density and temperature profiles. They
used a basic model of the heat transport inside the mantle to compute the
total planetary energy balance and calculate the relevant quantities
required to describe the thermal evolution of planets between 1 and
4.8 $\ME$.  Assuming rapidly rotating planets, they also predicted the
evolution of the PMF intensity from the dynamo scaling laws originally
developed by \citet{Christensen06} and later improved by
\citet{Aubert09} and \citet{Christensen10}. In Table
\ref{tab:thermal-models-results} we summarize some of the results
presented in Fig. 8 of \citet{Gaidos10}.

In the \CTE model the interaction between the core thermal structure,
the existence and growth of a solid inner core, the mantle properties,
the onset of plate tectonics and the surface temperature, determine
the existence and survival of an intense PMF in SEs.  A robust upper
bound, close to 2 $\ME$ (low-mass SEs) to have strong protective PMF
is obtained with this model.  Planets with masses larger than this
limit will not cool enough to develop a solid inner core, a condition
that considerably boosts the magnetic field intensity in lighter and
cooler planets.  As a consequence, massive SEs ($M\gtrsim 2 \ME$)
develop a decaying weak magnetic field.  In this case the dynamo is
shut down early when the convective power falls below the condition
for dynamo action.  These results depend strongly on the existence or
not of mobile lids (plate tectonics, PT).

\subsection{The \MTE model}
\label{subsec:MTEmodel}

\cite{Tachinami10} (\MTE model) pays special attention to the role
that the mantle have in the extraction of heat from the core, taking
into account its particular rheological properties.  They use the
Mixing Length Theory modified for solid planets in order to compute
the energy transported by convection in the mantle and the energy
budget at the CMB for planets with masses between 0.1 and 10 $\ME$.
They assume a planet with a mantle-core mass ratio similar to the
Earth's.  The \MTE model includes parameters like the rheological
properties of the mantle (especially important is the activation
volume V$^*$ that determines the viscosity dependence on pressure and
temperature) and the temperature contrast $\Delta T_{CMB}$ in the
boundary layer between the convecting mantle and the CMB.  After
solving the mantle and core coupled thermal transport equations, they
compute, for different model configurations, the heat flux at the CMB
$F_{CMB}$, the heat conducted along the adiabat $F_{cond}$ at the CMB,
and the solid inner core radius $R_{ic}$ as a function of time.  To
predict the PMF intensity and its lifetime they estimate the
convective power as $Q_{conv}=(F_{CMB}-F_{cond})\times 4\pi R_c^2$
where $R_c$ is the radius of the outer core.  As in the case of the
\CTE model they assume rapidly rotating planets, i.e. planets with a
dynamo operating in the dipolar regime.

Two important new predictions arise from the \MTE model: (1) thermal
evolution is affected by the strong dependence on pressure and
temperature of mantle viscosity and (2) the intensity and lifetime of
the PMF would strongly depend on the initial temperature profile which
is characterized by the parameter $\Delta T_{CMB}$, 
the temperature contrast between the core and the lower mantle; it is a
property mainly determined by the early accretion and differentiation
history of the planet.  Larger values of the initial temperature
contrast $\Delta T_{CMB}$, created for example by a violent accrection
history, would favor the appearance of an intense and long-lived PMF.
They found that by doubling $\Delta T_{CMB}$ from 1000 K to 2000 K the
dynamo lifetime in massive SEs ($M>2\ME$) is almost one order of
magnitude larger.  It should be stressed that the \CTE model also
studied the effects of different thermal profiles by considering the
cases of non-habitable surface temperatures.  This would be the case
of highly irradiated planets (not considered in the \MTE model).  In
this case the conclusions drawn by both models about the effect of
different temperature profiles on the PMF properties, were essentially
the same.

\subsection{Comparison of the models}
\label{subsec:Comparison}

The results of the \CTE and \MTE models point broadly in the same
direction: SEs with masses less than approximately 2 $\ME$ seem to
have properties better suited for the development of strong and
long-lived PMFs.  The $\sim$ 2 $\ME$ threshold is the most robust
prediction of these models.  In the \CTE model the reason that limits
the capability of low-mass SEs to sustain a dynamo is the early
cooling and stratification of the core.  On the other hand, the \MTE
model predicts a strong effect of the pressure-dependent viscosity in
the thermal evolution of the planet, which limits the capacity of
mantle convection to extract heat efficiently from the core,
especially in the case of massive planets.

\medskip

We want to highlight that the evolution of the PMF properties
predicted by the \CTE and \MTE models were obtained for planets with
short, but not specified, periods of rotation, i.e. planets with
dynamos operating in the dipolar regime.  For rapidly rotating
dynamos, the reference dynamo models used by the \CTE and \MTE models
\citep{Christensen06, Olson06, Aubert09, Christensen09} predict that
the magnetic energy density and hence the magnetic field intensity,
depends weakly on the rotation rate.  The still open question regarding 
the effect that long periods
of rotation would have in the PMF properties predicted by these
models.  Particularly interesting is to study the effect that
rotation would have in the determination of dynamo regime.  It would
be equally interesting to estimate the value for the rotation rate
where the rapid rotation approximation could be used.  These questions
are particularly relevant in the case of tidally locked SEs as will be
discussed in section \ref{subsec:class}.

For the purpose of this work a limited subset of
model configurations included in the \CTE and \MTE models and
highlighted in Table \ref{tab:thermal-models-results} have been selected.  
The final goal
is to study the effect of rotation in the evolution of long-lived
protective PMF in habitable SEs\footnote{A solid planet is considered
  habitable if the teperature on its surface is enough to maintain
  liquid water on the planet surface (see \cite{Kasting93} and
  references there in).}.  The chosen
configurations must meet two criteria: (1) habitable surface
temperatures (288 K in the \CTE model, 300 K in the \MTE model) and
(2) long-lived and intense PMF.  In the CTE case the latter condition
requires the existence of plate tectonics.  Since we are interested in
computing upper bounds to surface PMF intensities, given the same set
of planetary properties (mass, composition, surface temperature, etc.)
the case where plate tectonics arises will be the best suited to set
these bounds.

For planets with 1 $\ME$ the configuration selected in the \CTE and
\MTE models must reproduce the intensity of the present Earth's PMF
and be consistent with the thermal evolution of the Earth, i.e. they
must predict correctly the time of inner core nucleation.

In order to understand the role of rotation in the PMF evolution in
SEs we have to take into account the dynamo properties that depend on
rotation rate.  To achieve this we shortly review the relationship
between the thermal state of the planet and the properties of the
dynamo in the following section.

\section{Properties of convection-driven dynamos}
\label{sec:dynamo-props}

\subsection{Scaling laws}
\label{subsec:scaling-laws}

It has been suggested from numerical dynamo experiments that the
global properties of convection-driven dynamos can be expressed in
terms of simple power-law functions of a modified Rayleigh number
\citep{Christensen06,Olson06,Aubert09,Christensen09}.  The scaling
relationships found by these works involves the convective power
$Q_{conv}$, the core geometry, i.e. the outer core radius and the
vertical height of the liquid core $D=R_{c}-R_{ic}$ ($R_{ic}$ is the
inner core radius), the rotation rate $\Omega$ and other properties of
the core.

The power-based scaling laws used in this work are expressed using the
recent parametrization by \citet{Aubert09} where the properties of the
dynamo are scaled in terms of the adimensional convective power
density $p$,

\beq{eq:p}
p={Q_{conv}\over{\Omega^3 D^2\bar{\rho}_cV}}
\eeq

Here $\bar{\rho}_c$ and $V$ are the average density and total volume
of the convecting region respectively.  The use of $p$ in the scaling laws, instead
of the mass anomaly related Rayleigh number $Ra_Q$, is justified from
physical and numerical grounds (see section 2.2 in \cite{Aubert09}).

Two basic adimensional quantities have been used to characterize the
global properties of a dynamo \citep{Christensen06,Christensen10}: the
{\it Lorentz number} $Lo$ and the {\it local Rossby number} $Ro_l$.
$Lo$ is the adimensional magnetic field strength and is defined by
Eq. (22) in \cite{Aubert09} as

\beq{eq:Lorentz}
Lo=\frac{B_{rms}}{\sqrt{\bar{\rho}_c\mu_o}\Omega D}
\eeq

where $B_{rms}=(1/V)\int B^2 dV$, is the rms amplitude of the magnetic
field inside the convecting shell with volume $V$, $\mu_o$ is the
magnetic permeability.  $Ro_l$ measures the ratio of inertial to
Coriolis forces and is defined as

\beq{eq:LocalRossby}
Ro_l=\frac{U_{rms}}{\Omega L}
\eeq

where $U_{rms}$ and $L\sim D/\bar{l}$ are the characteristic
convective velocity and length scale respectively ($\bar{l}$ is the
mean spherical harmonic degree of the kinetic energy spectrum).  By
definition $Ro_l$ will be large for vigorous convection, small
characteristic length scale or slowly rotating dynamos.  Conversely
rapid rotating dynamos, large characteristic length scale or a weak
convection will produce small values of $Ro_l$.

Numerical dynamo experiments covering a wide range of physical
properties and boundary conditions
\citep{Christensen06,Olson06,Aubert09,Christensen09,Christensen10}
have found the following scaling relationships for $Lo$ and $Ro_l$
(Eqs. (22) and (30) in \citealt{Aubert09}):

\beq{eq:Lo-scaling}
Lo=c_{Lo} f_{ohm}^{1/2} p^{1/3}
\eeq

\beq{eq:Roml-scaling}
Ro_{l}^{*}\equiv{{Ro_l}\over{(1+\chi)}}=c_{Rol}\;p^{1/2} E^{-1/3}
(Pr/Pm)^{1/5} 
\eeq

where $f_{ohm}$ is the fraction of the available convective power
converted to magnetic field and lost by ohmic dissipation, $\chi\equiv
R_{ic}/R_c$, $E=\nu/(\Omega D^2)$ is the (viscous) Ekman number ($\nu$
is the viscous diffusivity) and $Pr/Pm = \lambda/\kappa$ is the ratio
of the Prandtl to magnetic Prandtl number ($\lambda$ and $\kappa$ are
the magnetic and thermal diffusivities respectively).  
For simplicity we have approximated the scaling law exponents to the ratio of the
smallest integers as suggested by \citet{Olson06}.  We have also
introduced here the {\it modified local Rossby number} $\Rol$
following the suggestion by \citet{Aubert09}.

The values of the constants $c_{Lo}$ and $c_{Rol}$ are obtained by 
fitting the results of numerical dynamo experiments with different
boundary conditions.  It should be noted that although the scaling law
for $Lo$ in Eq. (\ref{eq:Lo-scaling}) has been generally tested only
for dipolar dynamos, recently \citet{Christensen10} has found that the
magnetic energy $E_m\sim Lo^2$ still follows a 2/3-scaling law for
multipolar dynamos.  In that case the constant $c_{Lo}$ fitting the
properties of multipolar dynamos is smaller than in the dipolar case
by a factor $\approx 0.6$ \citep{Christensen10}.

Particulary interesting is that the scaling law for $Lo$ provides a
simple way to scale $B_{rms}$ with $p$ irrespective of the dynamo
regime.  Using the definition in Eq. (\ref{eq:Lorentz}) and the
scaling law in Eq. (\ref{eq:Lo-scaling}) we can write down an
expression for the magnetic field intensity (see Eq. 22 in
\cite{Aubert09})

\beq{eq:Brms-scaling}
B_{rms}=c_B f_{ohm}^{1/2} (\bar{\rho}_c\mu_o)^{1/2} \Omega D p^{1/3}
\eeq

Given the fact that by definition $p\propto\Omega^{-1/3}$ (see
Eq. (\ref{eq:p})), we find that $B_{rms}$ is almost
independent\footnote{It should be recalled that the exact value of the 
exponent of $p$ in the scaling law for $Lo$ is close to 1/3 but not
  exactly equal to this value.} of rotation rate both in the dipolar
and multipolar regimes.  However, the dipolar component
of the field will depend on rotation rate through the local Rossby
number in the case of reversing dipolar and multipolar dynamos 
(see section \ref{sec:rotation-PMF}).

\subsection{Dynamo regimes}
\label{subsec:dynamo-regimes}

Dynamos could be broadly classified according to the power spectrum of
the magnetic field at CMB in two groups: dipolar dominated dynamos,
i.e. dynamos where the dipolar component dominate over other higher
order components, and multipolar dynamos that have a flatter spectrum
or a weak dipolar contribution.

Quantitatively the dipolarity of core dynamos is commonly measured
using the ratio $f_{dip}$ of the mean dipole field strength at the CMB, 
$\bar{B}_{dip}$, to the rms field strength summed up to the harmonic
degree 12 at the same surface $\bar{B}_{CMB}$,

\beq{eq:fdip}
f_{dip}=\frac{\bar{B}_{dip}}{\bar{B}_{CMB}}.
\eeq

It has been assumed that dipolar dominated dynamo have $f_{dip}>0.35$
\citep{Aubert09, Christensen10}. Conversely when $f_{dip}\lesssim0.35$
the dynamo is classified as multipolar.  Another useful quantity to
measure the degree of dipolarity is the relation $b_{dip}$ between the
rms strength of the field in the shell volume $B_{rms}$ and the
dipolar component at the CMB,

\beq{eq:bdip}
b_{dip}=\frac{B_{rms}}{\bar{B}_{dip}}.
\eeq

\begin{figure}[htp]
  \centering
  \includegraphics[width=0.70\textwidth]{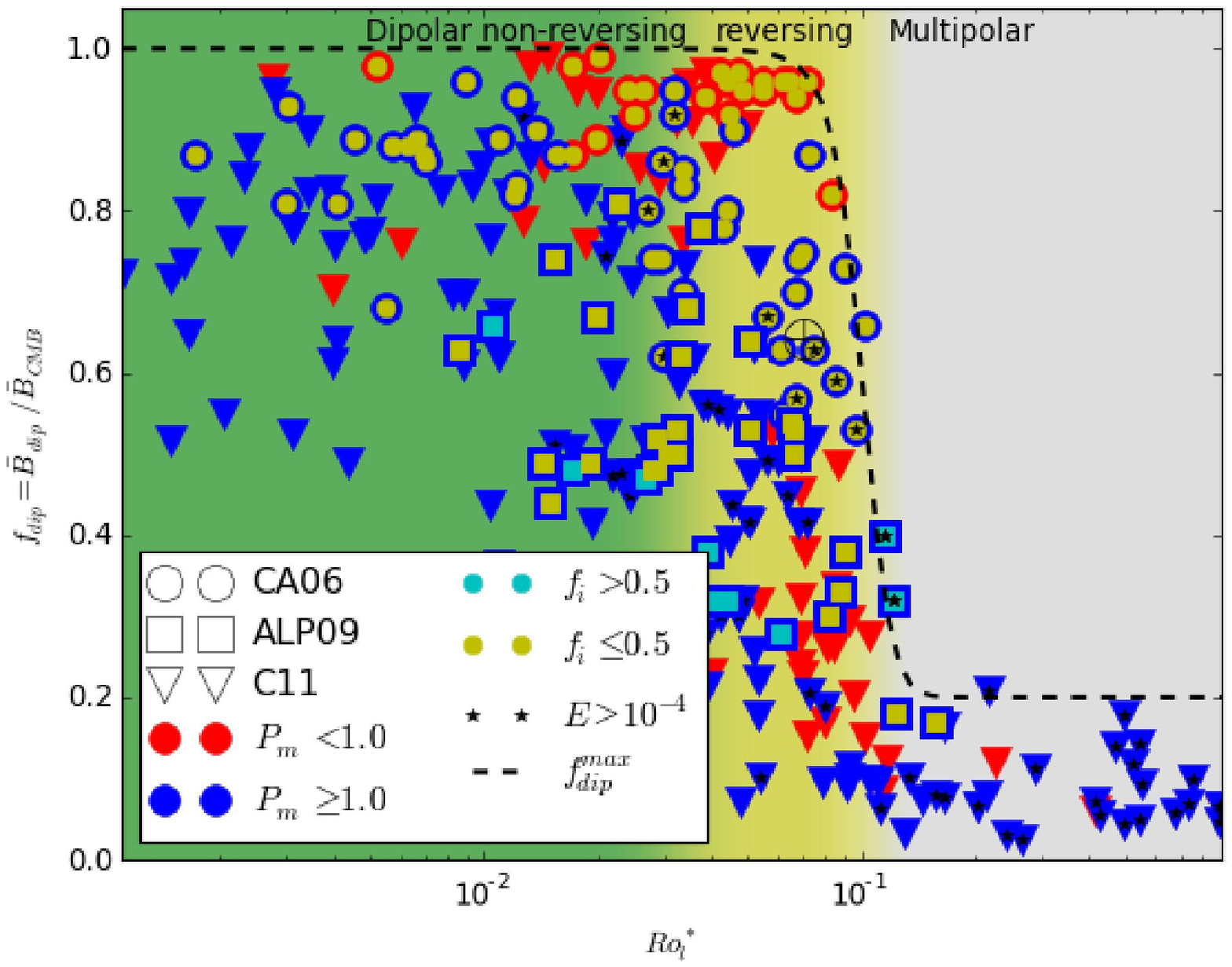}\\
  \includegraphics[width=0.70\textwidth]{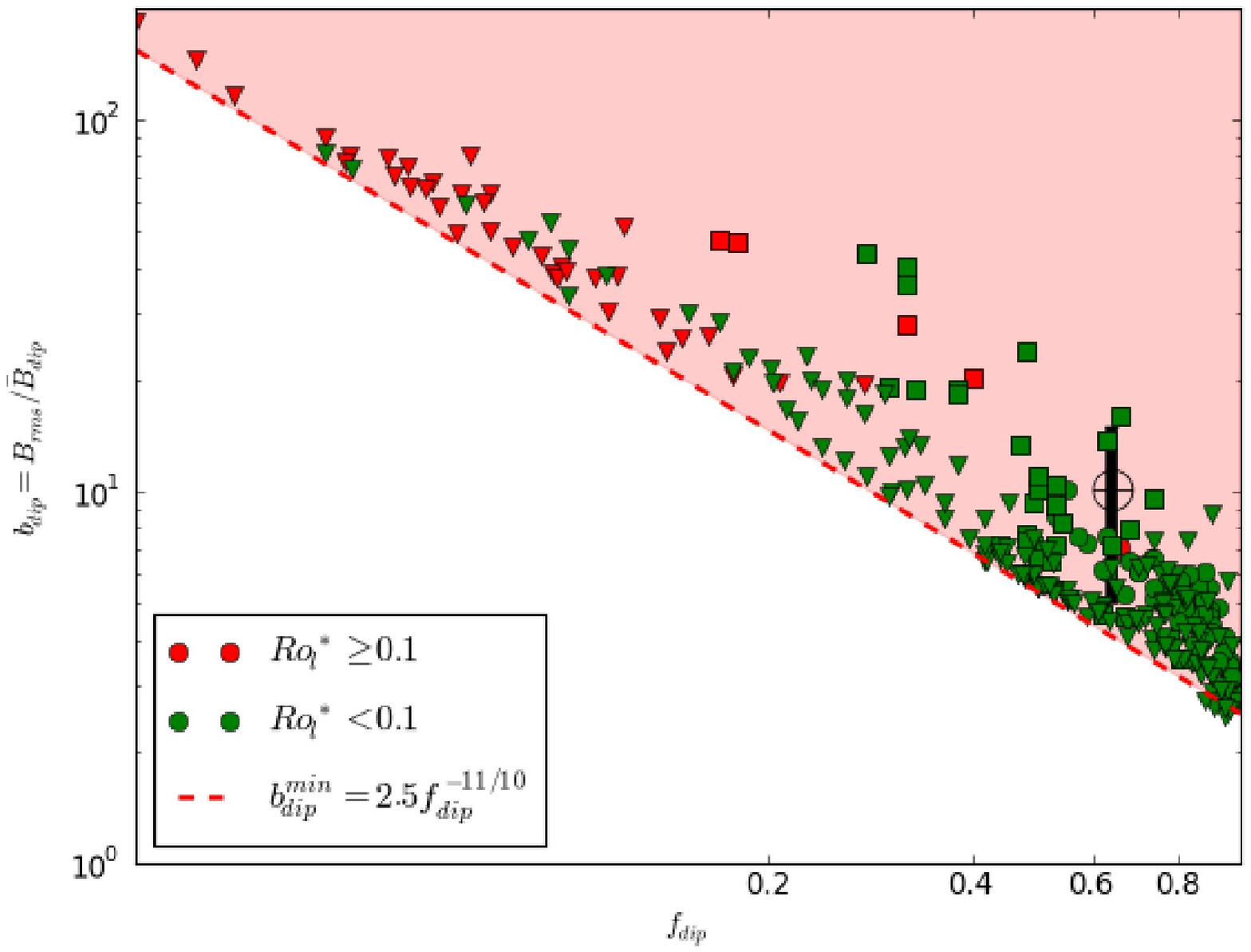}\\
  \caption{Upper panel: observed values of the $f_{dip}$ ratio in
    numerical dynamo experiments.  CA06, ALP09 and C11 stand for data
    obtained from \citet{Christensen06}, \citet{Aubert09} and
    \citet{Christensen11} respectively.  The dashed curve is the
    maximum value of $f_{dip}$ attainable at a given value of the
    local Rossby number.  The regions corresponding to different
    dynamo regimes have been schematically depicted using shading
    vertical bands.  Given the general complexity of the problem
    either multipolar dynamos could be found in the shaded region of
    reversing dipolar dynamos or non reversing dynamos ($f_{dip}\sim
    1$) with certain boundary conditions could be found in the
    reversing band.  Lower panel: values of $b_{dip}$ and $f_{dip}$
    for the same set of numerical dynamo results used in the upper
    panel.  A correlation between the minimum value of $b_{dip}$ at a
    given value of $f_{dip}$ is used in this work to estimate the
    maximum value of the dipolar component of the CMF (see text).  The
    position of the geodynamo is indicated with the symbol $\oplus$.
    In the lower panel the uncertainty in $b_{dip\oplus}$ is indicated with
    the thick black line.
  \label{fig:bdip-fdip}}
\end{figure}

Large values of $b_{dip}$ are typically a signature of a multipolar
dynamo although in numerical experiments dipolar dominated dynamos
could also have large $b_{dip}$ values.  However, the contrary is not
true: low values of $b_{dip}$ are only found in dipolar dominated
dynamos.  Typical values of $b_{dip}$ can be found in the lower panel
of Fig. (\ref{fig:bdip-fdip}).

For the present conditions of the geodynamo $f_{dip\oplus}\simeq 0.63$
($B_{dip\oplus}=0.263$ mT and $B_{CMB\oplus}=0.42$ mT
\citep{Olson07a}) placing it in the dipolar regime.  However, the
value of $b_{dip}$ for the geodynamo is largely uncertain.  Estimates
of $B_{rms}$ (required to compute $b_{dip}$) can be computed applying
proper scaling laws.  Using for example the Elsasser number criterion
an estimate of $B_{rms\oplus}\sim 4$ mT is obtained
\citep{Roberts00, Olson07a}.  Dividing by $B_{dip\oplus}=0.263$ we
obtain $b_{dip\oplus}\sim 15$.  On the other hand power based scaling
laws as those used in this work \citep{Christensen10} could be used to
predict $B_{rms}\sim 1.5$ mT (see e.g. \citealt{Aubert09}) and
therefore $b_{dip\oplus}\sim 5$.  Both values are also compatible
with those found in numerical dipolar dominated dynamos (see lower
panel of Fig. \ref{fig:bdip-fdip}).

Finally dynamos could be stable in the long-term or exhibit a
reversing behavior (see \cite{Amit10} and references there in).  In
most cases dipolar dynamos do not reverse and reversing dynamos are
multipolar, with very little (if any) overlap between dipolar and
reversing dynamos \citep{Kutzner02}, although some special dynamos are
dipole-dominated reversing (e.g. \citealt{Olson07}). Dipolarity and
reversals define the ``dynamo regime''.

Studying numerical dynamos in a wide range of conditions,
\citet{Christensen06} discovered that the local Rossby number is a
proxy of dynamo regime (see Fig. \ref{fig:bdip-fdip}). Their finding
has been confirmed by other works \citep{Olson06,Aubert09,Driscoll09}
over a wider range of dynamo parameters.  \citet{Aubert09} 
found that dipolar dominated magnetic fields are generated by dynamos
with values of $\Rol$ below a threshold of around 0.1.
\citet{Driscoll09} have confirmed this result but using as a
dipolarity proxy the modified Rayleigh number $Ra_Q$ (see Figs. 2 and
3 in \cite{Driscoll09}).  It should be stressed that although
$\Rol$ could not be the only controlling factor of dipolarity, most of
the available evidence points to this quantity as a good proxy for
dynamo regime.

The value of $Ro_l$ for the geodynamo is estimated at 0.08
\citep{Olson06}. For $\chi_\oplus=0.35$ the value of $\Rol=0.07$.
This places our planet close to the boundary between dipolar and
multipolar dynamos.  The reason the geodynamo is so close to that
particular boundary is unknown (see section \ref{sec:discussion} for a
discussion).

Dynamo regimes are thus separated in parameter space by complex
boundaries broadly limited by approximate values of $\Rol$.
Non-reversing dipolar dynamos has $Ro^*_l<0.04$ (irrespective of the
type of convection and boundary conditions).  The critical value of
$\Rol$ for the transition from dipolar to multipolar regime is in the
range $0.04<\Rol<0.1$ depending for example of the type of convection.
In numerical dynamo experiments, where several types of convection are
considered, this region is populated by a mixture of non-reversing,
reversing dynamos and multipolar dynamos.  We denote this interval in
$\Rol$ the ``reversing region''. Finally, dynamos with $\Rol>0.1$ are
multipolar (see upper panel Fig. \ref{fig:bdip-fdip}).

\section{The role of rotation in the PMF properties}
\label{sec:rotation-PMF}

Using the results of the thermal evolution models and the scaling laws
presented in section \ref{subsec:scaling-laws}, we could try to
constrain the expected properties of the PMF in SEs.  \citet{Gaidos10}
and \citet{Tachinami10} performed this task assuming rapidly rotating
planets and hence dipolar dominated dynamos.  This work go further
by including the effect of rotation into determining the dynamo
regime.  This section will develop a general procedure to
estimate the PMF properties in planets with dipolar and multipolar
CMF, a condition required to estimate the magnetic properties of
slowly rotating planets.

\subsection{Rotation and CMF regime}
\label{subsec:CMF-regimes}

In order to predict the evolution of CMF regime and to constrain the
evolving PMF intensity, is necessary to find a general expression for the
local Rossby number as a function of time, planetary mass and rotation
rate.  Replacing $p$ as defined in Eq. (\ref{eq:p}) and using
$E=\nu/(\Omega D^2)$, the scaling law for $Ro^*_l$ in
Eq. (\ref{eq:Roml-scaling}) can be written as:

\beq{eq:Roml-development}
Ro^*_l=c_{Rol}\times \left[Q_{conv}^{1/2}\Omega^{-3/2} D^{-1} (\bar{\rho}_c
V)^{-1/2}\right]\times\left[\nu^{-1/3} \Omega^{1/3}
D^{2/3}\right]\times(\lambda/\kappa)^{1/5}
\eeq

In this expression $Q_{conv}$ and $\chi=R_{ic}/R_c$ are provided
directly by the thermal evolution model.  Using $\chi$ is possible to compute
$D=R_c(1-\chi)$ and $V=(4\pi/3)R_c^3(1-\chi^3)$.  The core
radius and average density are scaled using $R_c\propto M_p^{0.271}$ and
$\bar{\rho}_c\propto M_p^{0.243}$ following \citet{Valencia06}.
Finally, the rotation rate $\Omega$ is expressed in terms of the
period of rotation $P$ using $\Omega=2\pi/P$ (see section
\ref{subsec:rotation} for further comments).  The general expression
for $Ro^*_l$ as a function of time, mass and rotation is finally

\beq{eq:Roml-MtP} 
Ro^*_l(t,M,P) = C \left[\bar{\rho}_c^{-1/6} R_c^{-11/6}\right]
\times \left[Q_{conv}^{1/2} (1-\chi)^{-1/3}
  (1-\chi^3)^{-1/2}\right] \times P^{7/6} 
\eeq

Here we have separated the quantities that explicitly depend on
planetary mass (first bracket) and those that depend explicitly on
time and implicitly on mass (second bracket).  The dependence on
rotation has been isolated in the $P^{7/6}$ term.  The quantity $C$
depends on the core viscous, thermal and magnetic diffusivities that
we will assume are nearly constant in time and almost independent of
planetary mass.  In our models $C$ has been set imposing that a 1$\ME$
planet at $t=4.54 \ Gyr$ and $P=1$ day has a value $\Rol(t=4.54\,Gyrs,
M=1\ME,P=1\,day)=0.07$.

As can be seen in equation (\ref{eq:Roml-MtP}) the value of $Ro^*_l$
strongly depends on the rotation period $P$. Slowly rotating dynamos
(large $P$) will have large values of $Ro^*_l$ and hence will be
multipolar.  It is also interesting to note that dynamos arising from
a completely liquid iron core ($\chi=0$) will have a smaller local
Rossby number and hence will be dipolar dominated for a wider range of
periods of rotation.  In this case, however, in the absence of
compositional convection the magnetic field strength could be much
weaker.  Additionally, for a fixed convective power, more massive
planets will have smaller values $\Rol$ and they tend to have dipolar
dominated dynamos.

\subsection{Scaling the magnetic field intensity}
\label{subsec:field-intensity}

Now the problem of estimating the CMF intensity in
the dipolar and multipolar regimes is tackled starting from the thermal evolution
model inputs $Q_{conv}$ and $D=R_c(1-\chi)$ (see
\ref{subsec:scaling-laws} for a $D$ definition).  Using the definition
of $b_{dip}$ (Eq. (\ref{eq:bdip})) and the scaling law for $B_{rms}$
(Eq. (\ref{eq:Brms-scaling})) the dipolar field intensity at the CMB
could be computed as,

\begin{eqnarray}
\label{eq:BdipBrms}
\nonumber\bar{B}_{dip} & = & \frac{1}{b_{dip}}c_B f_{ohm}^{1/2}
\sqrt{\bar{\rho}_c\mu_o} \Omega D p^{1/3} \\ & = &
\frac{1}{b_{dip}}c_B \sqrt{f_{ohm}\mu_o} \times
\left[\bar{\rho}_c^{1/6}
  R_c^{-2/3}\right]\times\left[Q_{conv}^{1/3}(1-\chi)^{1/3}(1-\chi^3)^{-1/3}\right]
\end{eqnarray}

Here the definition of $p$ (Eq. (\ref{eq:p})) was used and replaced 
$D$ and $V$ in terms of $R_c$ and $\chi$.  It is customary to
assume that in the case of rapid rotating dynamos $b_{dip}\sim 1$ and
hence $\bar{B}_{dip}\sim B_{rms}$.  That was the approximation used by
\cite{Gaidos10} and \cite{Tachinami10} to estimate the field
intensities reported in their works.  However, in many relevant cases
(see section \ref{subsec:application}) this approximation will not be
valid and a proper prescription to estimate $b_{dip}$ as a function of
the thermal evolution model is required.  We have studied the values
of $b_{dip}$ for a set of numerical dynamo results
\citep{Christensen06,Aubert09, Christensen10, Christensen11} and found
that although the value of this quantity varies in a wide range, both
in the case of dipolar and multipolar dynamos, it is possible to find
a lower bound $b^{min}_{dip}$ that is a function of $f_{dip}$ (see
Fig. \ref{fig:bdip-fdip}),

\beq{eq:bdip-min}
b_{dip}^{min}=c_{bdip} f_{dip}^{-\alpha},
\eeq

where $c_{bdip}\simeq 2.5$ and $\alpha\simeq 11/10$ are the best fitting
parameters.  Using this equation an upper bound to the dipolar field
at the CMB could be computed directly from Eq. (\ref{eq:BdipBrms}),
replacing $b_{dip}$ as $b^{min}_{dip}$ from Eq. (\ref{eq:bdip-min}),

\beq{eq:Bdipmax}
\bar{B}_{dip}\lesssim \bar{B}_{dip}^{max} = \frac{c_{Lo}}{c_{bdip}}
f_{dip}^{11/10} \sqrt{f_{ohm}\mu_o} \times \left[\bar{\rho}_c^{1/6}
  R_c^{-2/3}\right]\times\left[Q_{conv}^{1/3}(1-\chi)^{1/3}(1-\chi^3)^{-1/3}\right]
\eeq

The constant $c_{Lo}$ adopts different values according to the dynamo
regime: in the multipolar case ($f_{dip}\lesssim0.35$), $c_{Lo}\simeq
1$ and in the dipolar case ($f_{dip} > 0.35$), $c_{Lo}\simeq0.6$.
Therefore, $c_{Lo}$ depends implicitly on $f_{dip}$.

In summary, although $f_{dip}$ and $b_{dip}$ do not have unique values
for dynamos in a given regime, they are constrained from above and
below respectively, providing an interesting opportunity to constrain
the dipolar component of the CMF.  Estimating the local Rossby number
of a planetary dynamo with a given rotation rate it is possible to compute the
maximum value of $f_{dip}$ attainable by dynamos at that $\Rol$ using
numerical dynamo results (upper panel in Fig. \ref{fig:bdip-fdip}).
With this quantity in hand and using equation (\ref{eq:Bdipmax}) it is
possible to calculate the maximum value of the dipolar component of the
CMF.

It is possible to write down this procedure in the form of a simple algorithm. 
Given a planet with mass $M_p$ and period of rotation $P$, the maximum
dipolar component of the CMF at time $t$ could be computed using the
following procedure:

\begin{enumerate}

\item Using the thermal evolution model find $Q_{conv}(t)$ and
  $\chi(t)$.

\item Compute $Ro^*_l$ using Eq. (\ref{eq:Roml-MtP}).

\item Using $\Rol$ compute the maximum value of $f_{dip}$ as given by
  an envelop to numerical dynamo results in the $f_{dip}-\Rol$ space
  (see dashed line in in the upper panel of Fig. \ref{fig:bdip-fdip}).
  
\item Using $f_{dip}$ compute the maximum dipolar component of the
  CMF, $\bar{B}_{dip}^{max}$ using Eq. (\ref{eq:Bdipmax}).  Values
  adopted for $c_{Lo}$ are $\approx 1$ for dipolar dynamos
  ($f_{dip}>0.35$) and $\approx 0.6$ for multipolar dynamos
  ($f_{dip}<0.35$) \citep{Aubert09,Christensen10}.

\end{enumerate}

We are assuming $f_{ohm}\approx 1$ which is consistent with the goal
to obtain an upper bound to the CMF.

Using the maximum intensity of the dipolar component of the CMF and
assuming a low-conductivity mantle it is possible to estimate the
magnetic field at the planetary surface.  In the case of a dipolar
dominated CMF the surface field intensity scales simply as
$(R_c/R_p)^3$.  But, if the CMF has a more complex power spectrum, the
properties of the magnetic field at the surface are harder to
estimate.  However, since we are interested in the protective
properties of the PMF and they mainly depend on the dipolar component
of the field \citep{Stadelmann10}, we can still compute the maximum
dipolar component of the PMF, $\bar{B}_{s,dip}^{max}$, using the formula:

\beq{eq:Bdipsurf}
\bar{B}_{s,dip}^{max}=\bar{B}_{dip}^{max}\left(\frac{R_c}{R_p}\right)^{3}
\eeq

\ref{sec:appendix} shows that this result is valid irrespective
of dynamo regime.

\subsection{Evolution of rotation rate}
\label{subsec:rotation}

As was said in section \ref{subsec:dynamo-regimes}, the dependence on rotation 
was expressed in terms of the period $P$
instead of the rotation rate $\Omega=2\pi/P$.  $P$ is better suited to
study cases where rotation and orbital periods are related (tidal
locking) or in cases where large tidal effects from planet-stellar
interaction or due to hypothetical moons introduce simple long-term
variation of $P$ \citep{Varga98}.

The long-term evolution of rotation periods in terrestrial planets is
a complex subject that depends on many different effects ranging from
dynamical conditions at formation; catastrophic impacts; interior
processes changing the distribution of matter to tidal interactions
with the central star; close orbiting bodies or other bodies in the
planetary system (see \cite{VanHoolst09} and references therein).

To model the long-term variations of the rotation period, it is necessary to 
considered two simple extreme scenarios: (1) Constant period of
rotation $P_o$; this will be the case for tidally locked planets and
also for planets that have preserved their primordial rotation,
e.g. Mars. (2) Linearly increasing rotation period. This would be the
case for planets affected by strong tidal damping from the central
star, a close big moon or other planetary system bodies.  In the
latter scenario, and following the models used to study the long-term
variation of the Earth's period of rotation \citep{Varga98}, we have
assumed a simple linear variation,

\beq{eq:Pt}
P(t)=P_o+\dot{P_o}(t-t_o).
\eeq  

Here $P_o$ and $\dot{P_o}$ are respectively the rotation period at
$t_o$ and its rate of variation.  For an Earth-like rotation we
assumed $t_o=4.54$ Gyrs, $P_o=24$ h and $\dot{P}_o\approx 1.5$ h
Gyr$^{-1}$, values which are compatible with a primordial rotation of
$P(t=0)=P_{ini}\approx 17$ h \citep{Varga98, Denis11}.

\section{Results}
\label{sec:results}

We have computed the evolution of $\Rol$ using Eq. (\ref{eq:Roml-MtP})
and the values of $Q_{conv}(t)$ and $\chi(t)$ provided by the \CTE and
\MTE models in the cases highlighted in Table
\ref{tab:thermal-models-results}.  The results for different planetary
masses assuming $P_o$=24 h, both in the case of constant and variable
period of rotation, are depicted in Fig. \ref{fig:Roml-t-Po24}.

There are important differences between the results obtained with the
\CTE and the \MTE models.  These differences arise from the magnitude
and evolution of the convective power density in both models (see
Fig. \ref{fig:p-t}).  While in the \CTE model the available power
comes directly from the entropy dissipation inside the core, in the
\MTE model the amount of energy available for convection is simply
bounded by the energy extracted by the mantle through the CMB.  As a
consequence, the power density is almost one order of magnitude larger
and grows faster with planetary mass in the \MTE case than in the \CTE
model.  On the other hand, the energy flux through the CMB falls more
rapidly in the \MTE model than the entropy dissipated inside the core
in the \CTE model. This effect produces a net decrease in the power
density at later times, at least in the case of low-mass SEs.

The most noticeable change in the evolution of the $\Rol$ happens
during the so-called ``$p$-rebound'' just after the start of the inner
core nucleation at time $t_{ic}$.  In the \CTE model, $p$ goes through
a sudden and strong increase at $t_{ic}$ due to the combined effect of
a convective power increase, which is the result of the release of
latent heat and light elements, and the reduction in the vertical
height $D$ of the convecting shell.  The $p$-rebound in the \MTE model
is milder and comes mainly from the reduction in $D$.  In the \MTE
model the CMB flux $F_{CMB}$ that determines the estimated value of
$Q_{conv}$, is not sensitive to the new sources of entropy dissipation
in the recently formed inner core.  Although we are comparing both
models in equivalent conditions, it is clear that the \MTE model lacks
very important details that reduce the likelihood of the conclusions
drawn from the application of this model in our case.

Given these fundamental differences we will present separate analysis
of the predicted properties of the CMF for the \CTE and \MTE models in
the following paragraphs.

\subsection{Role of rotation in the \CTE model}
\label{subsec:CTEresults}

\begin{figure}[htp]
  \centering
  \includegraphics[width=0.70\textwidth]{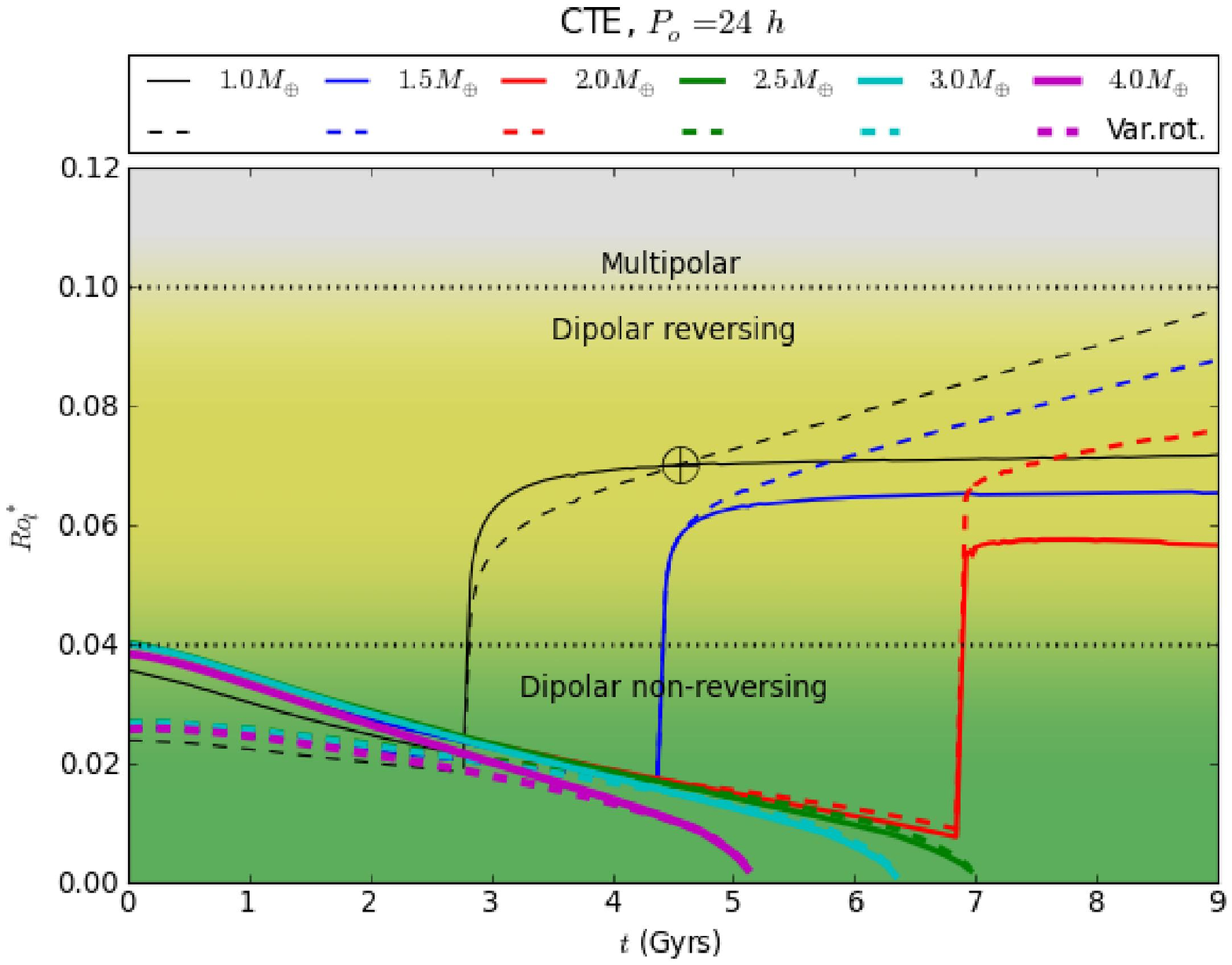}\\
  \includegraphics[width=0.70\textwidth]{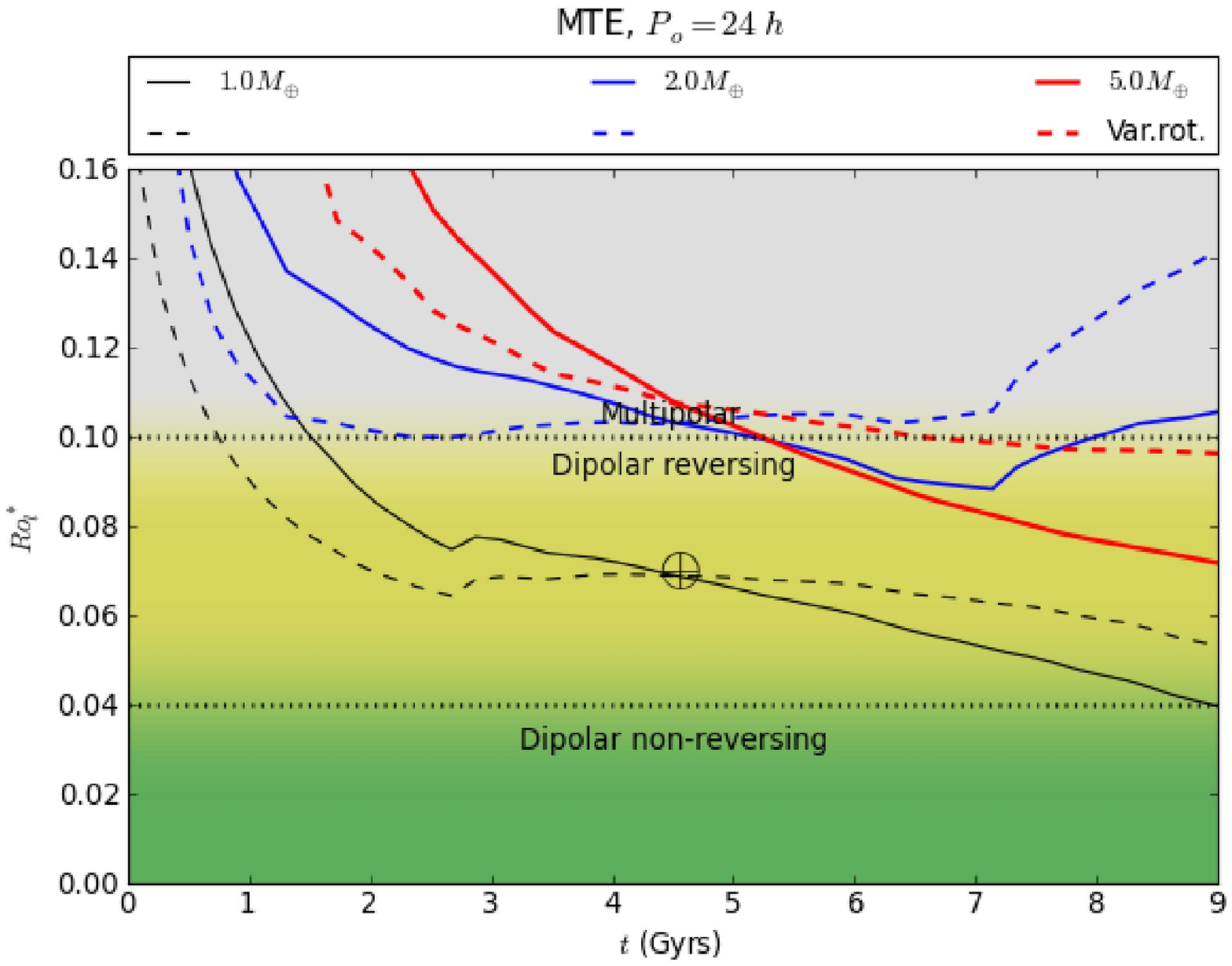}
  \caption{Evolution of the local Rossby number computed from selected
    results of the \CTE (upper panel) and \MTE (lower panel) models.
    Solid and dashed lines correspond to the cases of constant and
    variable period of rotation respectively.  Earth-like values for
    the period of rotation were assumed, i.e. $P_o$=24 h and $t_o$ =
    4.54 Gyrs.  In the case of a variable period of rotation (dashed
    lines) we have used $\dot{P_o}$ = 1.5 h Gyrs$^{-1}$.  Shaded
    regions enclose values of $Ro^*_l$ corresponding to multipolar
    (gray upper region), dipolar reversing (yellow middle region) and
    dipolar non-reversing (green lower region) dynamo regimes.
  \label{fig:Roml-t-Po24}}
\end{figure}

\begin{figure}[htp]
  \centering
  \includegraphics[width=0.8\textwidth]{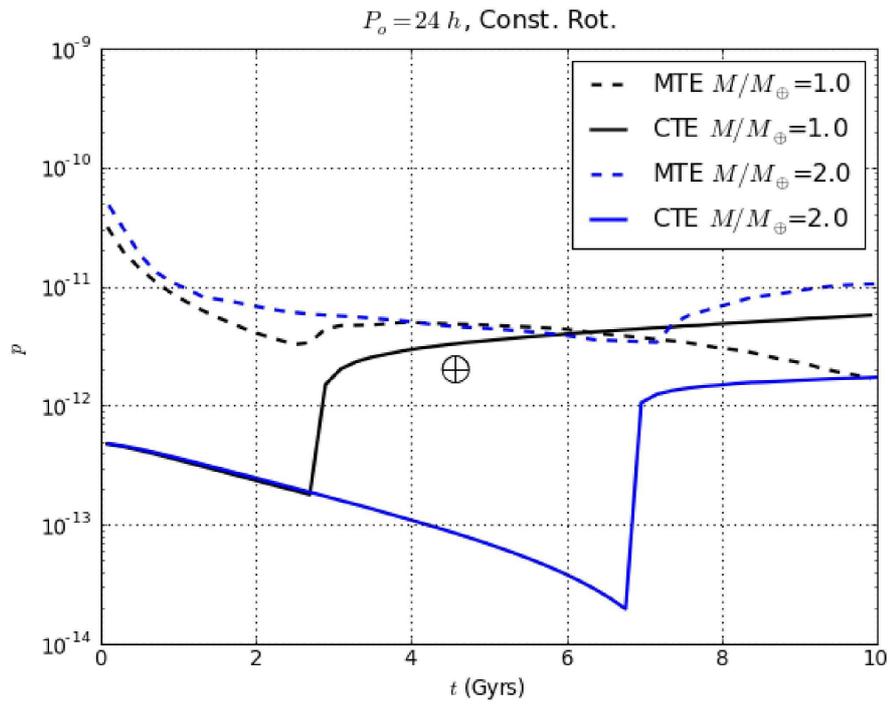}
  \caption{Evolution of the convective power density $p$ in the \CTE
    (solid lines) and \MTE (dashed lines) models for two different
    planetary masses.  $p$ is one order of magnitude larger in the
    \MTE model and falls faster than in the \CTE model due to
    differences in the estimations of the convective power.
  \label{fig:p-t}}
\end{figure}

The evolution of $\Rol$ for the \CTE model is depicted in the upper
panel of Fig. \ref{fig:Roml-t-Po24}.  As expected, the local Rossby
number varies in time in a similar way as $p$ does (see Fig.
\ref{fig:p-t}).  Evolution of the period of rotation has an
important effect, particularly when the convective power flux has
dropped at late times.

In the case of low-mass planets, the $p$-rebound is responsible for
the most important features of the $\Rol$ evolution.  The sudden
increase in $p$ and the related decrease in $D$ produce an even faster
increase in $Ro^*_l$.  This behavior has two effects: (1) $Ro^*_l$
reaches a minimum value at $t_{ic}$, e.g. $Ro^{*}_{l,min}\simeq0.02$
for $M=1\ME$ and $Ro^{*}_{l,min}\simeq0.01$ for $M=2\ME$, regardless
of variations in the period of rotation, and (2) the inner core
nucleation, at certain periods of rotation (see below) will mark the
transition from a dipolar to a multipolar regime.

In the case of massive planets where the solid inner core does not
appear in the first 10 Gyrs, $Ro^*_l$ steadily decreases until the
dynamo finally shuts down.  In this case the ability of the planet to
create a dipolar CMF at large periods of rotation (slowly rotating
planets) is constrained by the value of the local Rossby number close
to the time of dynamo shut down.  We found in this case that
$Ro^*_{l,min}\sim 10^{-3}$, which is again independent of variations
in the period of rotation.

Taking into account that $Ro^*_l\propto P^{7/6}$
(Eq. (\ref{eq:Roml-MtP})) and using the critical value $Ro^*_{l}=0.1$,
at which all dynamos have became multipolar, we found that planets
with a constant period of rotation $P\lesssim
P_{mul}=P_o(0.1/Ro^*_{l,min})^{6/7}$ will have a chance to develop a
dipolar dominated CMF at some point in the dynamo lifetime.  In the
case of planets with $M\lesssim 2 \ME$ the limit is $P_{mul}\sim 4-9$
days (the lowest value corresponding to the lightest planet, 1 $\ME$ 
and for more massive planets, i.e. $M\gtrsim 2 \ME$, $P_{mul}\sim
30-50$ days.  These limits set upper bounds of what should be
considered rapidly rotating planets, i.e. planets able to develop
strong and long-lived dipolar magnetic fields.  To find more precise
limits it is necessary to compute the dipolar CMF lifetime, a problem that
will be addressed below.

\begin{figure}[htp]
  \centering
  \includegraphics[width=0.8\textwidth]{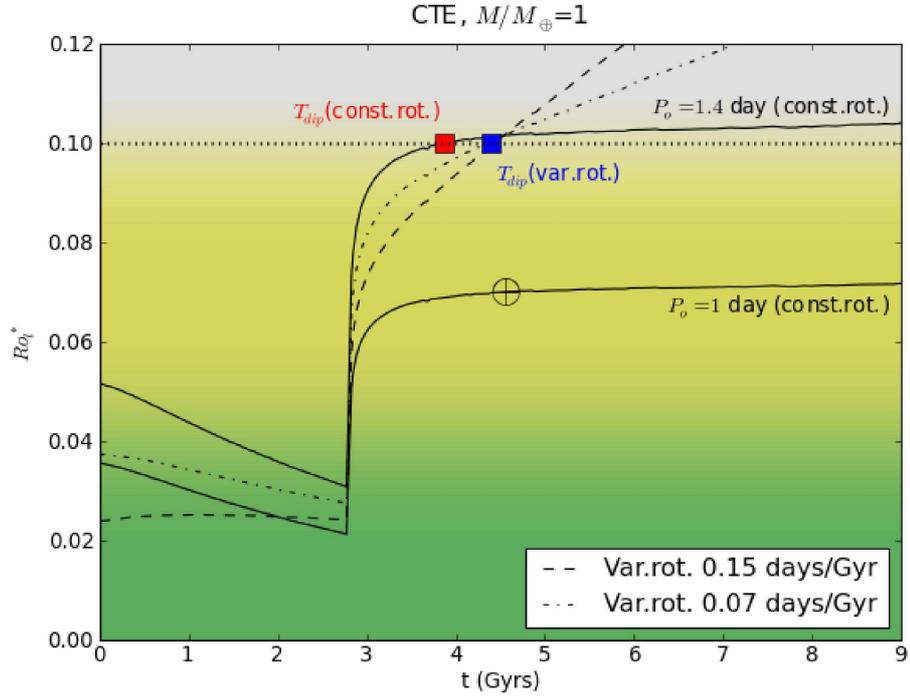}
  \caption{Comparison of the evolution of $Ro^*_l$ for a 1$\ME$ planet
    with a period of rotation $P\approx 1.4$ days and $P\approx 1$ day
    in the \CTE model.  The $Ro^*_l$ in the case of variable period of
    rotation (dashed and dashed-dotted lines) has been computed for
    two different rates $\dot{P_o}$ assuming the same reference period
    of rotation $P_o\approx 1.4$ days at $t=4.54$ Gyrs.  The squares
    are placed at the times where the dynamo becomes multipolar
    (dotted line at $\Rol=0.1$ marks the transition to the multipolar
    regime) for the constant and variable period of rotation cases
    respectively. Notice that the dashed line is below the
    dashed-dotted line before $t=4.54$ Gyrs and above it after that
    time.  Shaded regions enclose values of Rol* corresponding to
    different dynamo regimes, multipolar (upper region, $\Rol>0.1$),
    dipolar reversing (middle region, $0.04<\Rol<0.1$) and dipolar
    non-reversing (lower region, $\Rol<0.04$).
  \label{fig:Roml-t-Po33}}
\end{figure}

In order to understand the role of rotation in the survival of a
dipolar dominated CMF in the case of planets with $M<2\ME$, the
evolution of $Ro^*_l$ must be studied for several values of the
reference period of rotation $P_o$ in the case of constant and
variable rotation rate.  The case for a $1 \ME$ planet with
$P_o\approx1.4$ days (33 h) is shown in Fig.
\ref{fig:Roml-t-Po33}. In this case the CMF dipolarity is guaranteed
only during the first few Gyrs when no solid inner core has been
formed and the convective power density is still down (low $Ro^*_l$).
Shortly after the $p$-rebound the convective forces increase enough to
make the CMF multipolar and the period of dipolar dominance ends.  The
time spent by the dynamo in the dipolar dominated regime is called the
{\it dipolar lifetime}, $T_{dip}$.  More massive SEs $M>2\ME$ exhibit
a different behavior.  For periods of rotation $P\lesssim P_{mul}$ the
CMF is multipolar at the beginning and turns dipolar at a time called
the dipolarity switch time, $t_{sw}$.  In this case $T_{dip}$ is
computed as the difference between the dynamo lifetime and the
$t_{sw}$.

\begin{figure}[htp]
  \centering
  \includegraphics[width=0.8\textwidth]{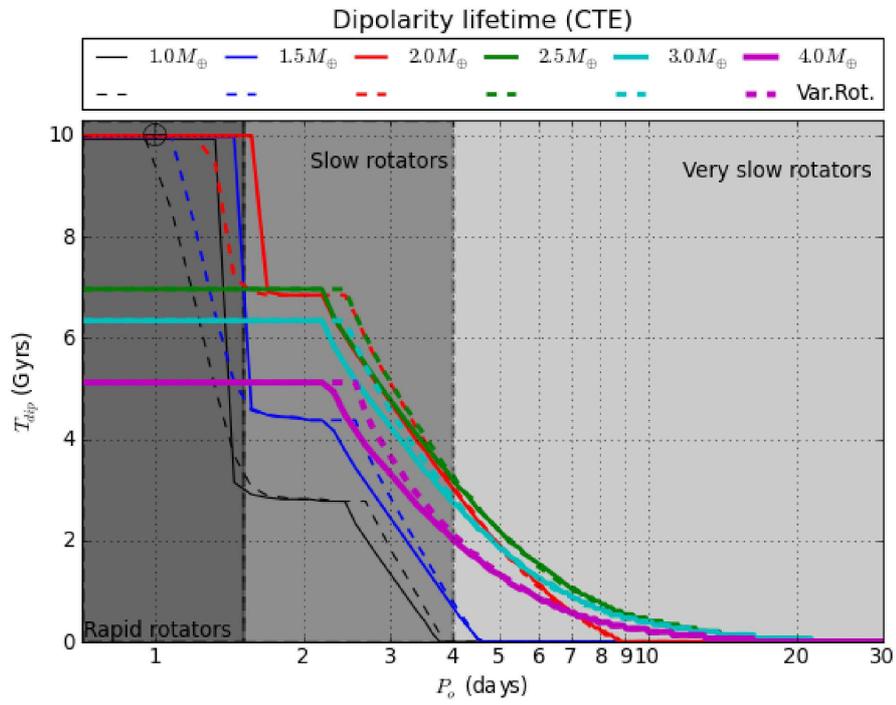}
  \caption{Lifetime of the dipolar dominated CMF in the \CTE model
    obtained from the analysis of the $Ro^*_l$ evolution for planets
    with different masses.  Constant (solid line) and variable (dashed
    lines) periods of rotation have been assumed.  The gray regions
    are limited by the maximum rotation periods of low mass planets
    ($M<2\ME$) in the three categories introduced in this work (see
    text).
  \label{fig:Tdip-Po-CTE}}
\end{figure}

A plot of $T_{dip}$ for different planetary mass as a function of the
reference period of rotation $P_o$ is depicted in Fig.
\ref{fig:Tdip-Po-CTE}.  As expected $T_{dip}$ is equal to the dynamo
lifetime in the case of rapidly rotating planets ($P_o\sim 1$ day).
The critical period of rotation $P_{c}$ up to which this condition is
met, increases with planetary mass for $M\leq2 \ME$ as a consequence
of a later formation of the inner core in more massive planets (see
Fig. \ref{fig:Roml-t-Po24}).  In the case of planets with $M>2\ME$,
the critical period $P_c$ is almost the same, $P_{c}\simeq 2-2.5$
days, a behavior explained by the similarity between the evolution of
$\Rol$ in this type of planets (see upper panel in Fig.
\ref{fig:Roml-t-Po24}).

A variable period of rotation decreases $P_c$ for planets below $2
\ME$ but increases in the case of more massive planets.  The
explanation of this behavior could be found in the upper panel of
Fig. \ref{fig:Roml-t-Po24}.  Low mass planets has a lower $P_c$ if
$\Rol$ is larger at late times and this is exactly the effect of a
variable period of rotation. Conversely more massive planets has a
larger $P_c$ if $\Rol$ is lower at early times as expected again in
the case of a variable period of rotation.

It is also interesting to note that in the case of low mass SEs, an
intermediate level of dipolar CMF lifetime beyond $P_c$ is found.
This intermediate level is a by product of the steep increase in
$Ro^*_l$ around the time of inner core nucleation $t_{ic}$.  For times
$t<t_{ic}$ the dipolarity condition is ensured in a wide range of
rotation rates.  This effect explains why the dipolar lifetime at this
intermediate level is close to $t_{ic}$. With these results it is possible 
to set a more stringent limit to the periods of rotation required to have
long-lived ($T_{dip}\gtrsim 3$ Gyrs) dipolar dominated CMFs.  We found
that regardless of the mass, SEs with $P\lesssim2-3$ days will meet
this condition.

An interesting and nontrivial prediction arises from the dipolar
lifetime dependence on mass and period.  In the works by
\citet{Gaidos10} and \citet{Tachinami10} low-mass planets were
identified as the best candidates to have intense, long-lived PMF.
This result was a consequence of the favorable conditions that the
early nucleation of a solid inner core and a lower viscosity mantle
have on the determination of the PMF properties and its lifetime.
This result is still valid here, at least for planets with periods of
rotation smaller than $\sim1.5$ days.  However, when rotation
periods are larger than this limit the roles are interchanged: massive
Super Earths would develop a dipolar CMF for a longer time than the
lightest planets.  The reason for this switch is the strong
$p$-rebound when the inner core starts to nucleate in light SEs.
Since the thermal evolution of massive planets does not exhibit such
rebound, the rotation period for $M>2\ME$ could be increased even
more, before the CMF becomes multipolar.

\begin{figure}[htp]
  \centering
  \includegraphics[width=0.7\textwidth]{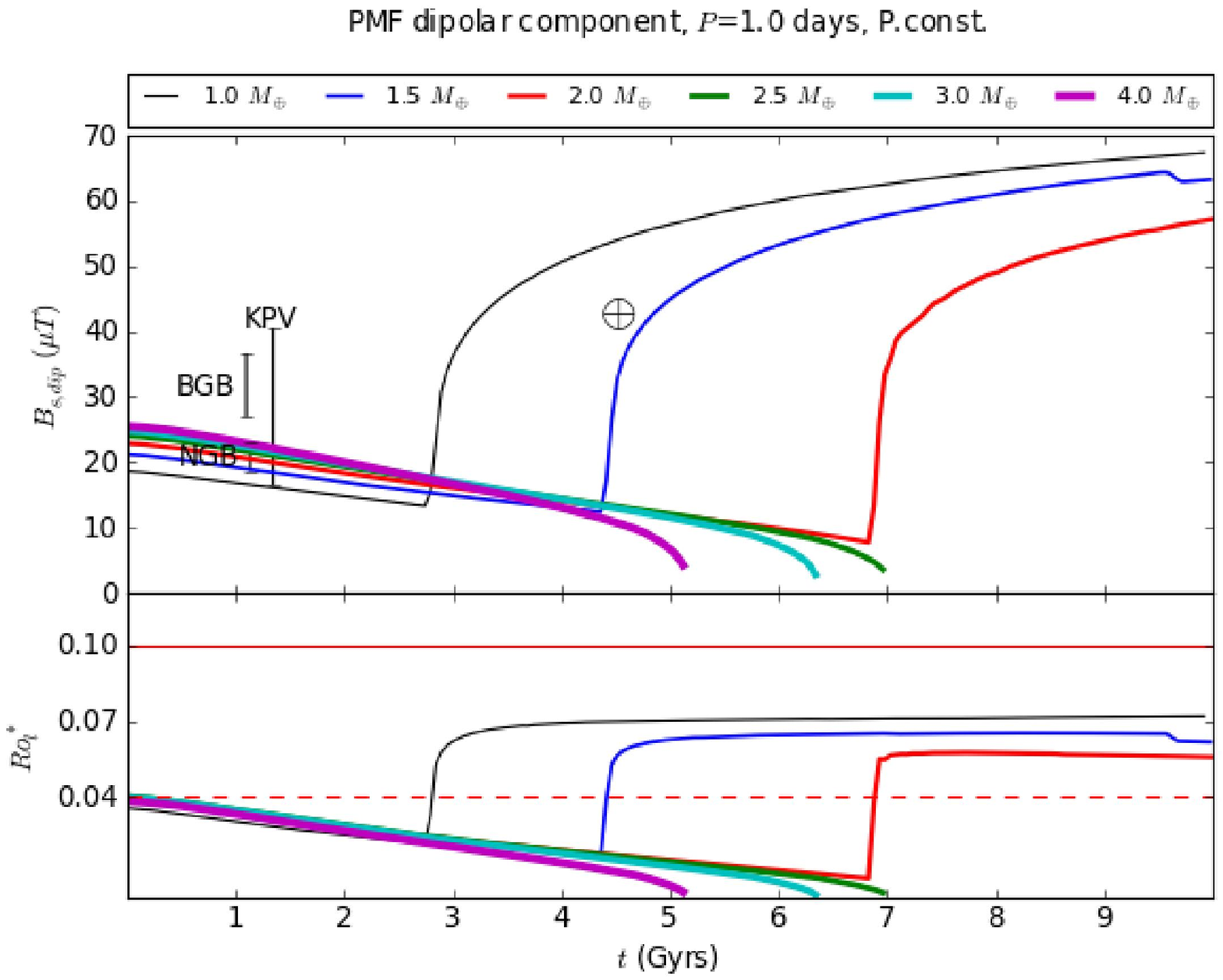}
  \includegraphics[width=0.7\textwidth]{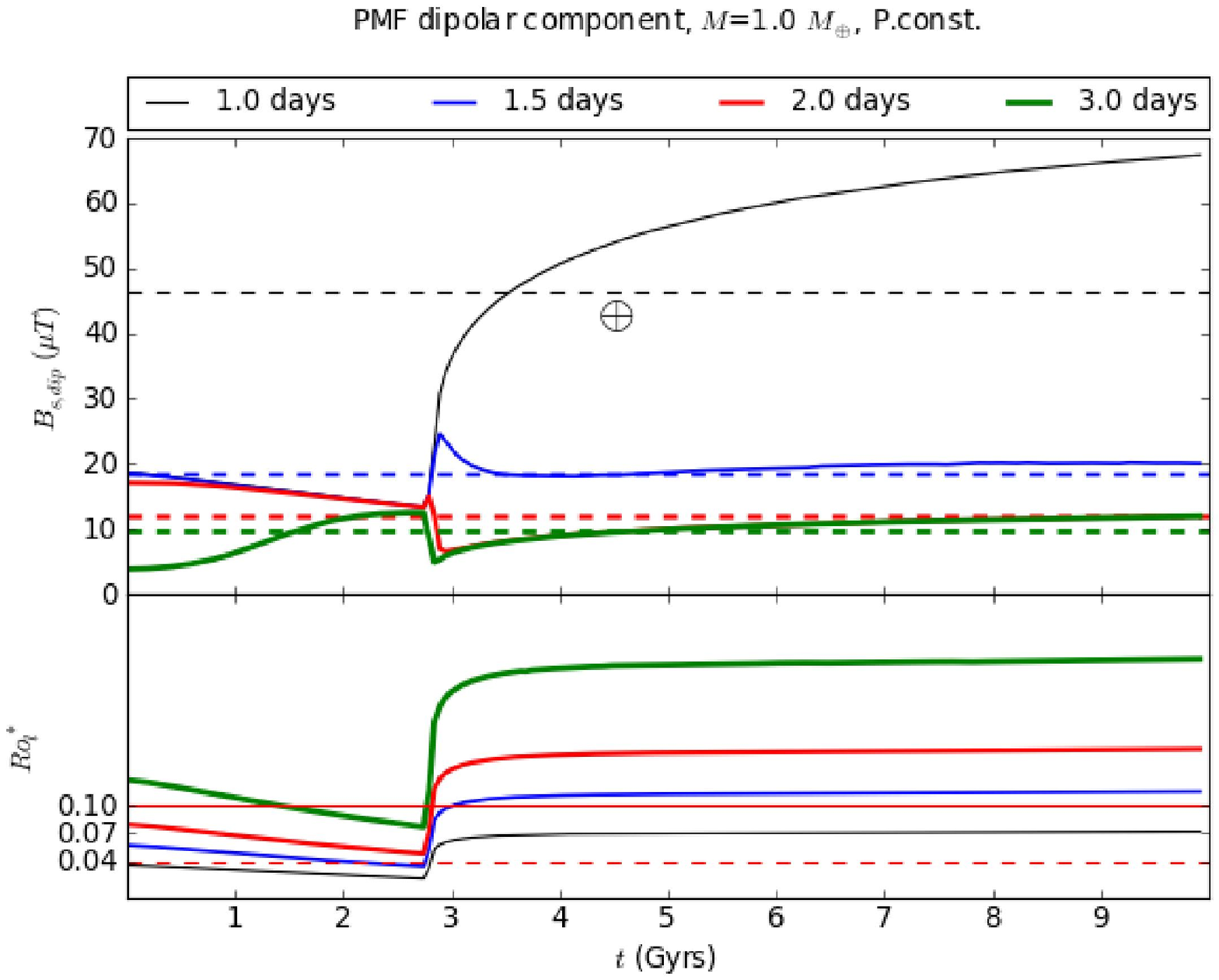}
  \caption{Maximum dipolar component of the surface PMF estimated with
    the procedure devised in this work and using the thermal inputs of
    the \CTE model.  In the upper panel the case of rapidly rotating
    planets, $P_o=1$ day, is presented. In the lower panel the dipolar
    PMF intensities for a $1 \ME$ planet with different periods of
    rotation are compared.  The value of the $Ro^*_l$ has also been
    plotted in order to illustrate the effect that a transition
    between dynamo regimes have in the evolution of the PMF.  The red
    continuous, dashed and dotted lines in the $\Rol$ subpanels are
    the limits between the regimes (dipolar non reversing, dipolar
    reversing and multipolar).  In the lower panel, the dashed lines
    in the magnetic field plot are the lifetime average of the maximum
    dipolar component of the PMF, $B_{avg}$ (Eq. (\ref{eq:Bavg})).
    The value of $B_{s,dip}$ for the present Earth's magnetic field
    (Earth symbol $\oplus$) and recent paleointensities measurements
    (error bars) has also been included (\citealt{Tarduno10}: Kaap
    Valley (KVP), Barberton Greenstone Belt (BGB), and Nondweni
    Greenstone Belt (NGB), dacite localities).
  \label{fig:Bs-t}}
\end{figure}

Taking into account the evolution of $Ro^*_l$ and the information it
could provide on the CMF regime, we have computed the intensity of the
dipolar component of the PMF, $\bar{B}_{s,dip}$ following the
procedure outlined in section \ref{subsec:field-intensity}.  The
result is depicted in Fig. \ref{fig:Bs-t} where we have additionally
included the measured value of the Earth magnetic field at present
times \citep{Maus05} and three recent measured paleomagnetic
intensities at 3.2 and 3.4 Gyrs ago \citep{Tarduno10} (indicated with
an Earth symbol and error bars respectively).  It is clear in this
case that the evolution of the $Ro^*_l$ has an important impact on the
measured field at the planetary surface.  In the case of an Earth mass
planet with $P=1.5$ days (lower panel in Fig. \ref{fig:Bs-t}) the
dipolar component of the field reaches a maximum intensity almost
$300$ Myrs after the initiation of the inner core nucleation just to
decay again in hundreds of megayears as a consequence of the increase
in intensity of the convective currents vs. the weak Coriolis
force.  At 3 Gyrs the CMF becomes fully multipolar and the intensity
measured at the surface is just the CMF dipolar component.  We have
assumed that the transition from the dipolar to the multipolar regime
happens in short times when compared with the thermal evolution time
scales.

In order to compare the dipolar PMF intensities of planets with
different masses and to understand the global role of rotation in the
determination of the PMF properties, the dipolar 
field intensity averaged over the dynamo lifetime $T_{dyn}$ was computed

\beq{eq:Bavg}
B_{avg}\equiv<\bar{B}_{s,dip}>=T_{dyn}^{-1}\int_0^{T_{dyn}}
\bar{B}_{s,dip} dt
\eeq

\begin{figure}[htp]
 \centering
 \includegraphics[width=0.8\textwidth]{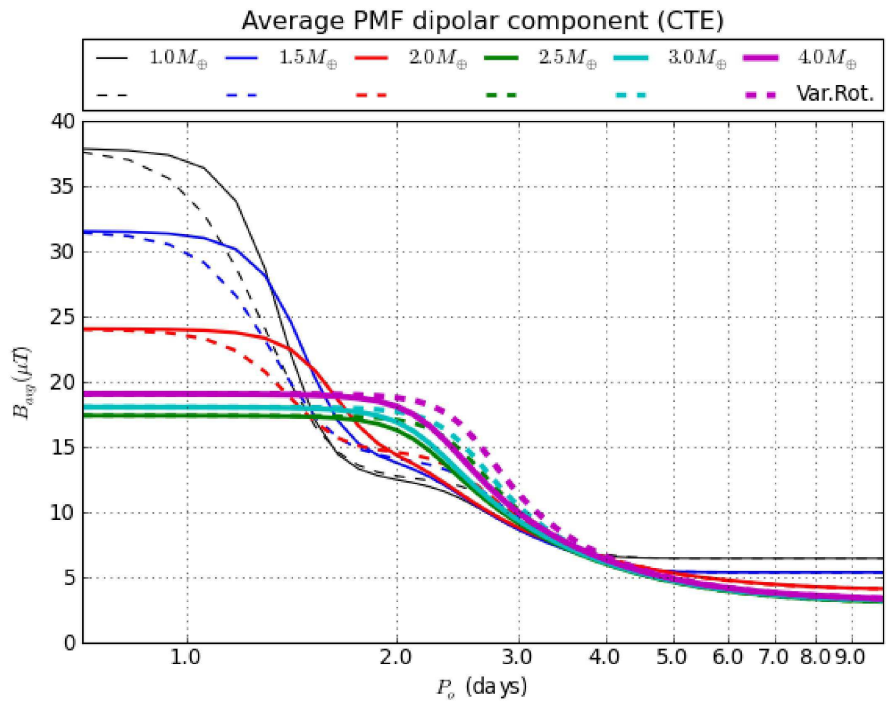}
 \caption{Surface dipolar field intensity averaged over the dynamo
  lifetime, $B_{avg}$ as defined by Eq. (\ref{eq:Bavg}), for planets with 
  constant (solid) and variable (dashed
  lines) periods of rotation.
 \label{fig:AvgBs-T}}
\end{figure}

Although not phenomenologically relevant this quantity is very useful
to compare the PMF for evolving dynamos with different planetary
parameters.  On the other hand the minimum and maximum ``historical''
dipolar field intensities for a given planetary mass and rotation
rate, observed in a relevant range of parameters, are of the same order of this
average and therefore $B_{avg}$ seems a
good first order estimate of the evolving PMF intensity.  In Fig.
\ref{fig:AvgBs-T} we plot $B_{avg}$ vs. $P_o$ for planets with
different masses.  We found a similar behavior of $B_{avg}$ as that
found in $T_{dip}$ (Fig. \ref{fig:Tdip-Po-CTE}).  In this case however
the differences between the field intensities are not as noticeable as
those found in the case of the dipolar CMF lifetime.

Field intensity is not the only observable we can try to estimate with
this model.  Reversal frequencies could also be inferred
\citep{Driscoll09}.  Although we have not computed this property it
should be mentioned that the predicted monotonous variation of the
local Rossby number could not explain completely the variation in
reversal frequency observed in the Earth's paleomagnetic field.  For
example the increase in $\Rol$ after inner core nucleation could
explain the transition from a superchrone to reversals but not from
reversals to superchrone.

\begin{figure}[htp]
 \centering
   \includegraphics[width=0.70\textwidth]{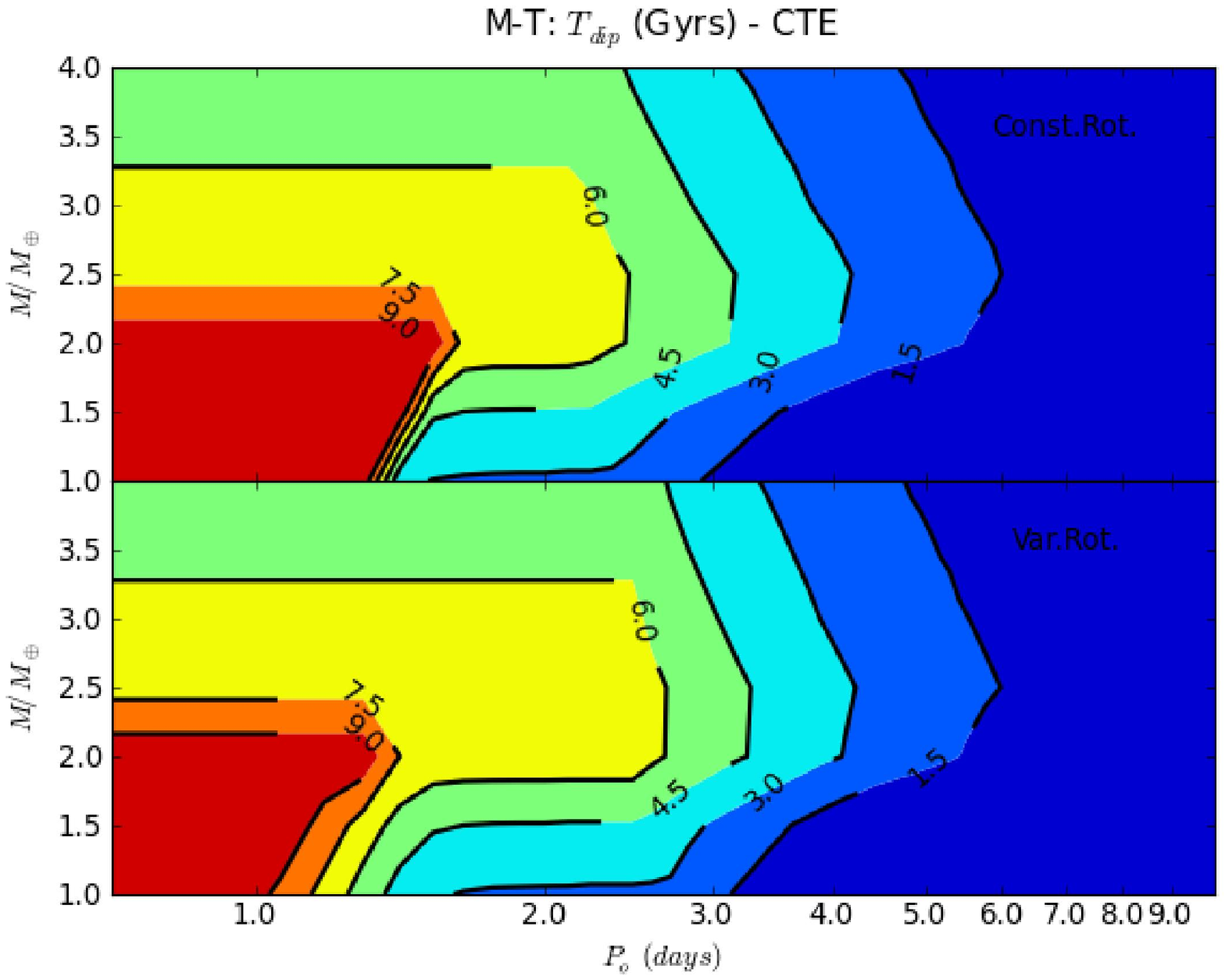}\\
   \includegraphics[width=0.70\textwidth]{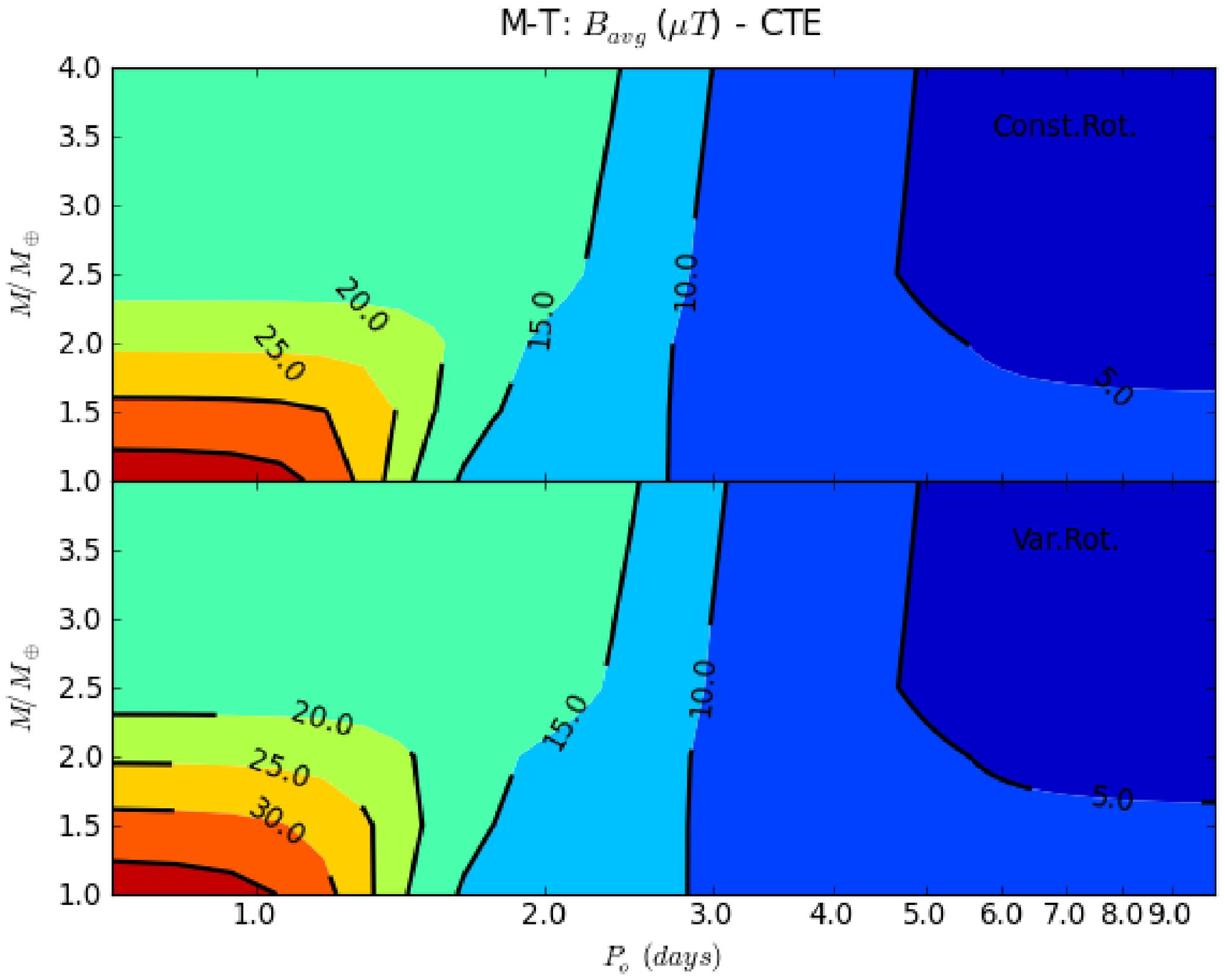}
   \caption{Mass-Rotation (\textit{M-P}) diagrams for the dipolar field
     lifetime, $T_{dip}$ and the average surface magnetic field,
     $B_{avg}$ in the \CTE model.  In each diagram the case for
     constant (upper half of each panel) and variable (lower half of
     each panel) periods of rotation have been assumed.  Contours of
     equal values of the quantities represented on each diagram are
     also included.
   \label{fig:MP-CTE}}
\end{figure}

Displaying the properties of an evolving magnetic field (dipolar
lifetime and intensity) for different planetary masses and periods of
rotation is challenging.  To simplify the graphical representation of
PMF properties and their dependence on mass and rotation rate, we
introduce here the Mass-Period diagram (hereafter \textit{M-P} diagram).  We
have used \textit{M-P} diagrams in Fig. \ref{fig:MP-CTE} to depict
$T_{dip}$ and $B_{avg}$ for the \CTE model, comparing the cases of
constant and variable periods of rotation.  The conclusions drawn from
Figs. \ref{fig:Tdip-Po-CTE} and \ref{fig:AvgBs-T} are well
illustrated in the \textit{M-P} diagrams (see Fig. \ref{fig:MP-CTE}).  
Planets with masses between 1 and 3 $\ME$ and periods
$P<3$ days, are the best suited to developing a long-lived dipolar
field, and planets with $M<2 \ME$ and $P<1.5$ days develop dipolar
field intensities of the same order as the Earth's.  Qualitatively
the effects of a variable period of rotation are negligible.

\subsection{Role of rotation in the \MTE model}
\label{subsec:MTE}

\begin{figure}[htp]
  \centering
  \includegraphics[width=0.70\textwidth]{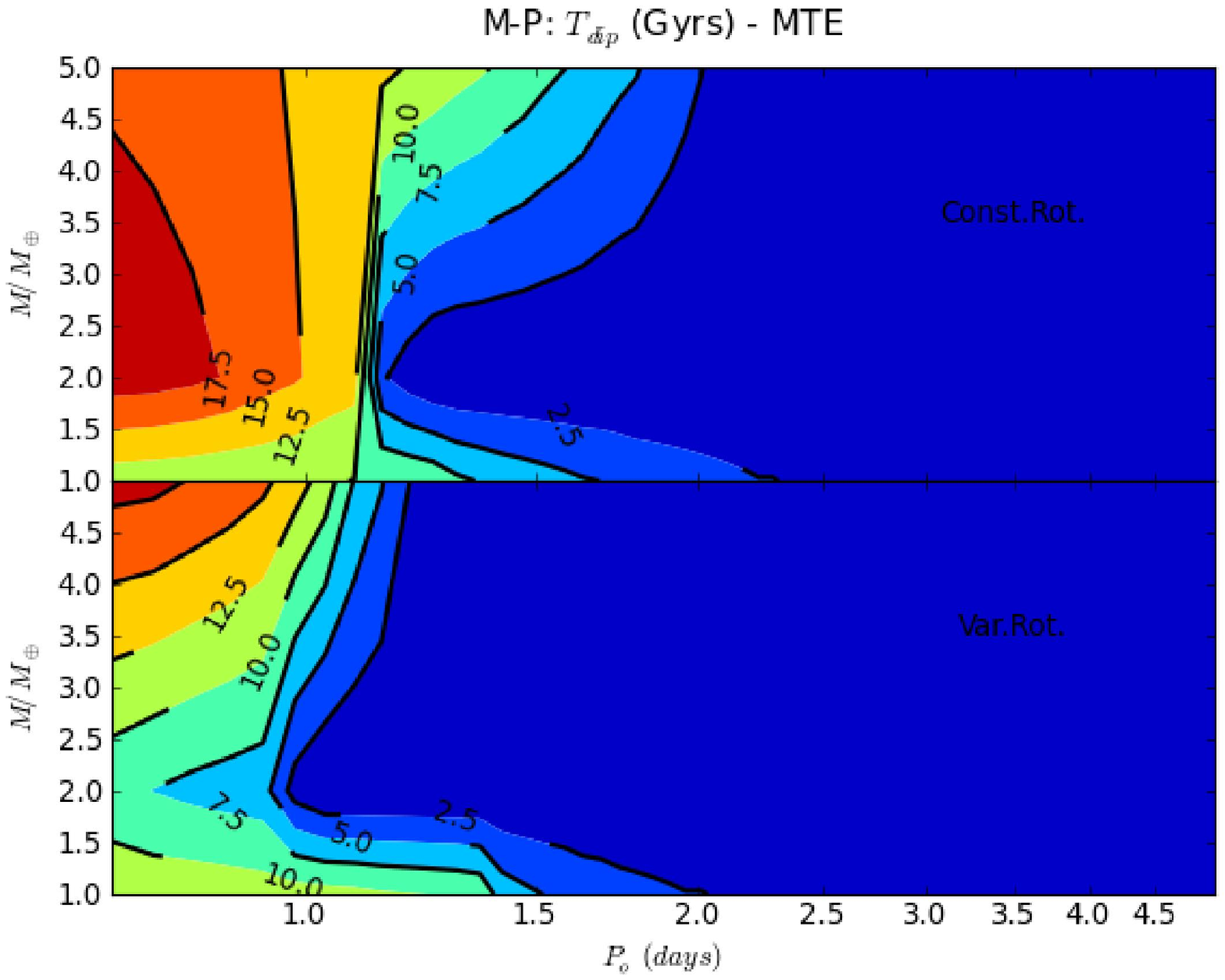}\\
  \includegraphics[width=0.70\textwidth]{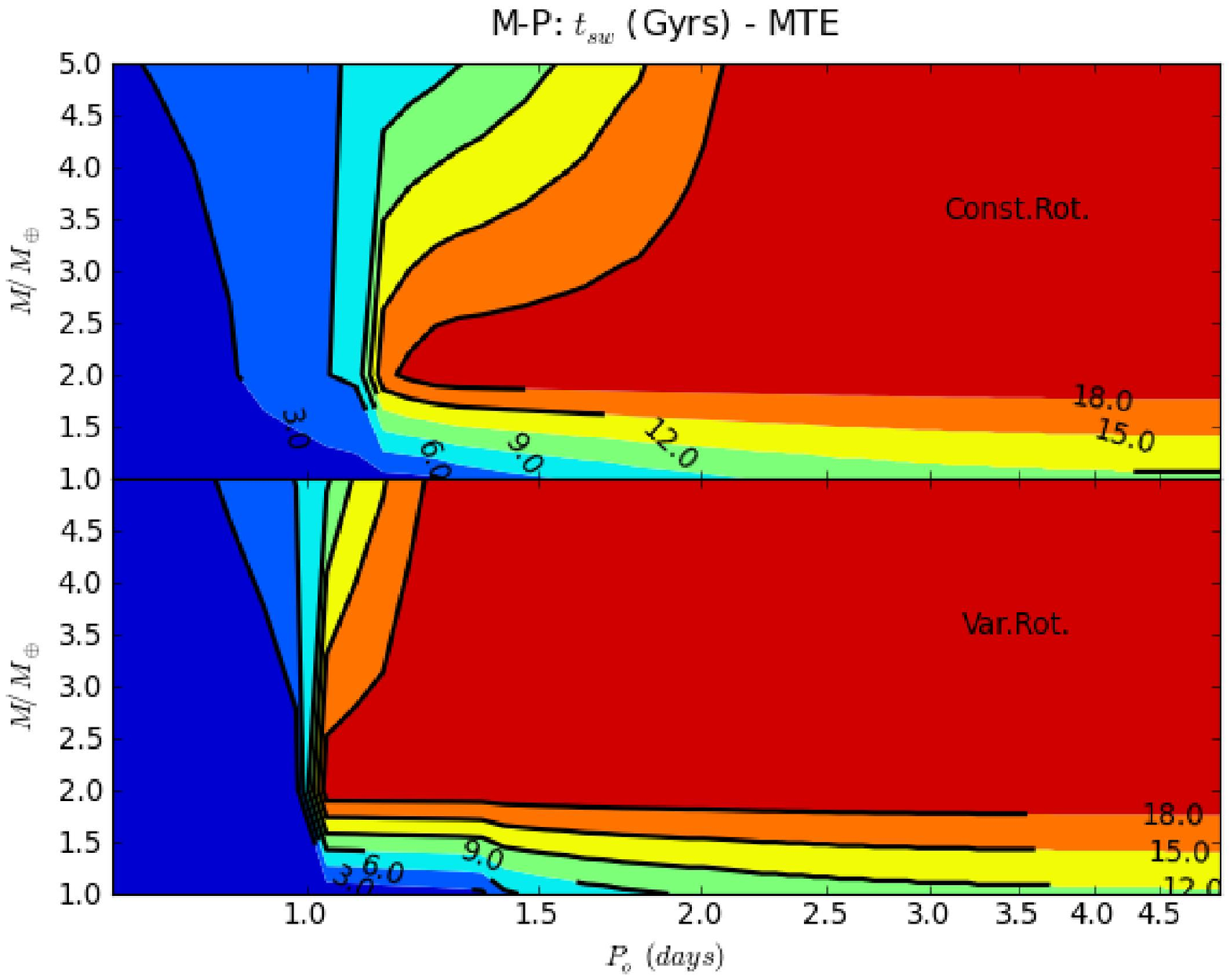}
  \caption{Mass-Rotation (\textit{M-P}) diagrams for the dipolar field
    lifetime, $T_{dip}$ and the dipolarity switch time $t_{sw}$ in the
    \MTE model.  In each diagram the case for constant (upper half of
    each panel) and variable (lower half of each panel) periods of
    rotation have been assumed.  Contours of equal values of the
    quantities represented on each diagram are also included.
  \label{fig:MP-MTE}}
\end{figure}

The evolution of $Ro^*_l$ for the \MTE model is plotted in the lower
panel of Fig. \ref{fig:Roml-t-Po24}. In all the cases the $Ro^*_l$
has large values at early times and hence the predicted CMF is always
multipolar at the beginning.  This result is a consequence of the
large initial CMB flux $F_{CMB}$ used in this model to estimate the
convective power at the core.  For all the planetary masses and in the
case of rapidly rotating planets, dynamos switch from a multipolar to
a dipolar regime at what we have called the dipolar switch time
$t_{sw}$.  In several cases, the intensity of the $p$-rebound and/or
the variation of the period of rotation reduce the time spent for the
dynamo in a dipolar dominated state $T_{dip}$.  In the analogous to
Fig. \ref{fig:MP-CTE} for the CTE model, we have plotted $t_{sw}$
and $T_{dip}$ in \textit{M-P} diagrams in Fig. \ref{fig:MP-MTE}.  In order to
have a protective PMF it is expected that the core CMF becomes dipolar
as early as possible (small $t_{sw}$).  Indeed a protective PMF is
more important in the early phases of planetary and stellar evolution
evolution than at late times (see e.g. \citealt{Lichtenegger10}).  We
also expect that the duration of the dipolarity phase $T_{dip}$ be
also large.  The effects of a variable period of rotation are
noticeable especially in the case of large mass planets.  This fact is
the result of the particular behavior of the evolving $Ro^*_l$.

For the \MTE model the limit for rapidly rotating planets, as estimated
by the condition to have long-lived dipolar fields, is more stringent
than in the \CTE model.  Only planets with periods shorter than 1 day
develop intense and long lived dipolar CMF in contrast with
approximately 2-3 days limit found in the case of the \CTE model.

\subsection{Rotation based SEs classification}
\label{subsec:class}

The resulting effects that a variation in the rotation period has in
the dynamo properties of SEs, suggest the possibility to classify SE
dynamos in three different groups: rapid, slow and very slow rotators
(see Fig. \ref{fig:Tdip-Po-CTE}).  {\bf Rapid rotators} are planets
with strong dipolar dominated dynamos lasting several to tens of
Gyrs. This type of planets have dynamos with local Rossby numbers
below the critical value, i.e. operating in the dipolar regime, for
the most part of the dynamo lifetime ($50\%$ is the criterion used
here).  Our planet would belong to this group.  {\bf Slow rotators}
are SEs with dipolar dynamos lasting for less than $50\%$ of the
dynamo lifetime.  Dynamos of this type spends most time of their
lifetime in the reversing region and a non negligible time above the
critical value for the multipolar regime.  {\bf Very slow rotators}
correspond to planets that do not develop a dipolar dominated field
during their entire dynamo lifetime.  Dynamos of this type have large
super critical local Rossby numbers.

The range of periods of rotation defining the proposed categories,
varies with mass and depends on the particular thermal evolution model
used to predict the PMF properties.  In the case of the \CTE model we
found that rapid rotators have $P\lesssim 1.5$ days, irrespective of
planetary mass, slow rotators with masses $M\lesssim 2 \ME$ have
periods of rotation in the range $4\lesssim P\lesssim 10$ days, while
more massive slow rotators will have $4\lesssim P\lesssim 20$ days.
Planets with $P\gtrsim 10-20$ days are very slow rotators. We have
plotted these limits in Fig. \ref{fig:Tdip-Po-CTE}.  In the case of
the \MTE model the limit for rapid rotators are more stringent than in
the \CTE model.  This fact is a consequence of the high convective
power predicted by this model at early times (see lower panel in
Figs. \ref{fig:Roml-t-Po24} and \ref{fig:p-t}).  The maximum
rotation rate for rapid rotating planets in this case is not larger
than $1$ day.  The category of slow rotators is practically nonexistence
in this model.  Planets with periods of rotation larger than $1-1.5$
days will be very slow rotators if the thermal evolution model has the
features predicted by the \MTE model.


\subsection{Application to already discovered SEs}
\label{subsec:application}

The application of these results to study the development of protective
magnetic fields in newly discovered SEs, depends on the ability to
know or estimate their rotation periods.  The most of already
discovered SEs are tidally locked and hence their rotation periods
are also known.  According to recent claims, planets with masses in
the range of interest for this work ($M<2 \ME$) and orbital periods
less than 50 days could be common around low-mass stars
\citep{Howard10}.  Therefore, tidally locked SEs are the best initial
target for this kind of analysis.  Nevertheless, in the case of
unlocked SEs yet to be discovered the feasibility to measure the
period of rotation using present and future observational facilities
has also been devised (see \cite{Ford01} and \cite{Palle08} and references
therein).

\begin{figure}[htp]
\centering
\includegraphics[width=0.9\textwidth]{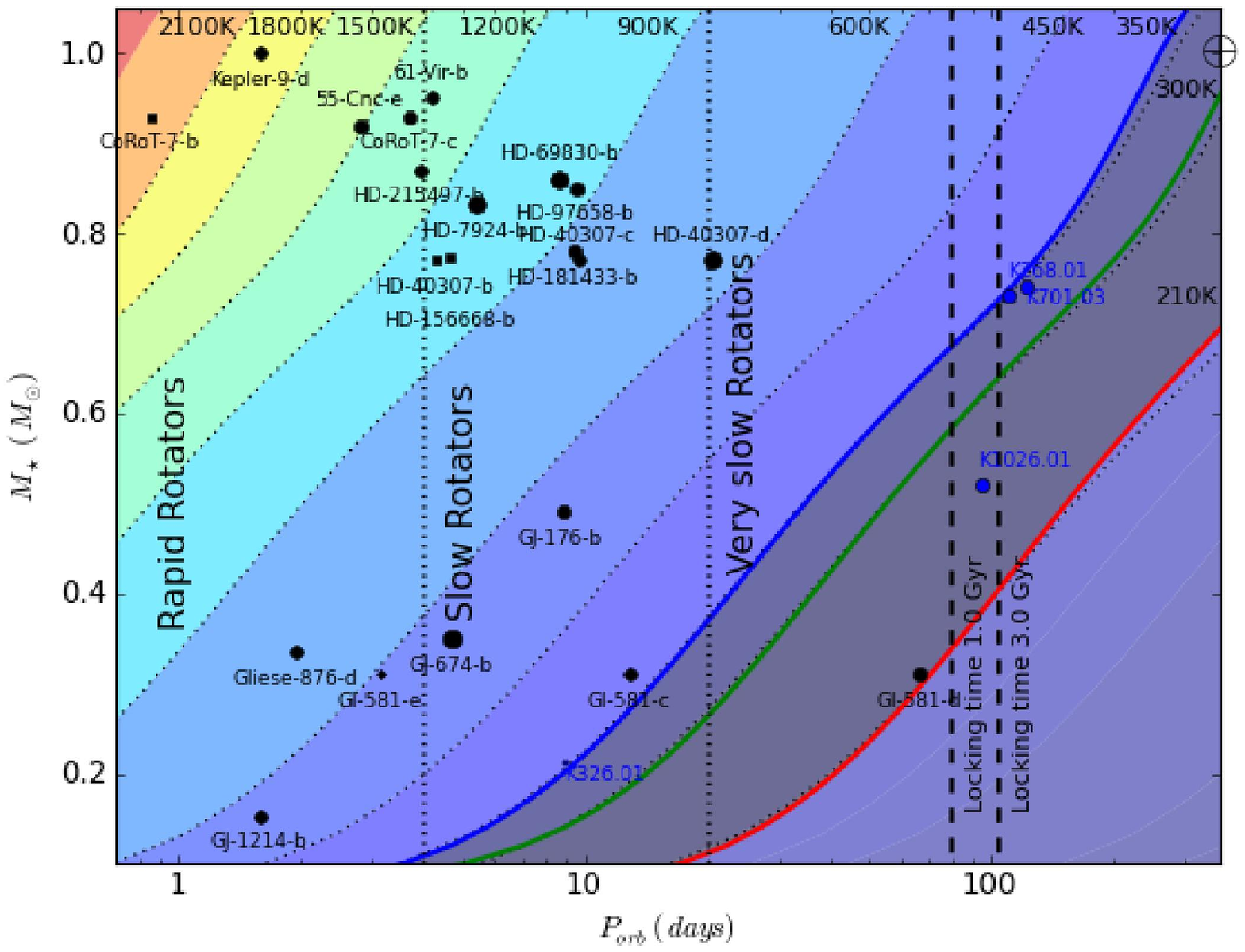}
\caption{Stellar mass vs. orbital period diagram for planets around
  GKM stars ($M<1.05 M_\odot$).  Circles indicate the position of 21
  of the known SEs and 4 Kepler candidates orbiting stars in the
  selected mass range.  Each circle has a diameter proportional to the
  minimum mass of the planet.  Contours of equal equilibrium
  temperatures $T_{eq}$ as measured in planetary surfaces assuming an
  Earth-like bond albedo ($A=0.29$) and a redistribution factor of 2
  as expected for tidally locked planet \citep{Selsis07a} has been
  also plotted.  Blue and red lines are the inner and outer limit of
  the habitable zone as computed using the Venus and Mars criteria in
  \citep{Selsis07a}.  Green line is the 1 AU-equivalent distance where
  the planet will receive the same flux as the Earth
  \citep{Kaltenegger10a}.  The vertical thick dashed line indicates
  the maximum distance inside which planets will be tidally locked in
  times less than 1 and 3 Gyrs
  \citep{Kasting93}.
\label{fig:CHZ-tidlock}}
\end{figure}

In Fig. \ref{fig:CHZ-tidlock} we have plotted the habitable zone
(HZ) for GKM stars and the physical parameters of confirmed SEs plus
the subset of Kepler candidates inside the HZ \citep{Borucki11}.  We
have also included there, contours of equilibrium temperatures
\citep{Selsis07a, Lammer10}, the outer limits to have tidally locked
planets in 1 and 3 Gyrs \citep{Kasting93}, and the range of periods
defining our categories of rapid, slow and very slow rotators in
the case of massive SEs assuming the \CTE thermal evolution model.

It is needed to point out that the results have a limited range of
applicability when compared with the wide range of physical properties
founded in this set of discovered SEs.  We have assumed from the
beginning that planets have a composition similar to Earth, habitable
temperatures and mobile lids.  Low density planets as GJ 1214b
probably covered by thick volatile atmospheres \citep{Nettelman11},
could have other ways to create intrinsic magnetic fields and the
limits to develop protective PMF should be different than those found
here in the case of core dynamos.  Nevertheless, rotation will still
play an interesting role in the PMF properties of these types of
planets and efforts to include this effect should not be
underestimated.  On the other hand the composition of many of these
SEs is still unknown (there is a lack of complete information on their
masses and radii).  SEs with high surface temperatures, e.g. Corot 7b
\citep{Valencia10}, also fall outside the range of applicability of
these results.  Higher surface temperatures reduce the viscosity in
the mantle favoring the extraction of energy from the core and
increasing the convective power energy.  \cite{Gaidos10} found higher
values of the surface PMF in the case of $\approx O(1) M_\oplus$
planets and larger dynamos lifetime in more massive SEs, when surface
temperatures are increased.  In those cases the $\Rol$ will also be
larger and the critical period of rotation to have multipolar dynamos
will be smaller.  These results suggest that hot SEs would lack of
protective magnetic fields even with periods of rotation proper of
colder slow rotators.

\section{Discussion}
\label{sec:discussion}

This work relies basically on three hypotheses: (1) the thermal
evolution models by \cite{Gaidos10} and \cite{Tachinami10}
provide global robust features of the thermal evolution in SEs, (2)
The scaling laws fitted with numerical dynamo experiments can be
extrapolated to regions in the parameter space where real planetary
dynamos lie, (3) the local Rossby number could be used as a proxy for
dynamo regime.

It is clear that to test hypothesis (1) it is necessary to address the open
questions left by \cite{Gaidos10} and \cite{Tachinami10}.  It is
important to solve a complete model including a rigorous treatment of
convection in the mantle, as was done in the \MTE model, but also
taking into account a detailed model of the structure and entropy
balance in the core as done by the \CTE model.  Despite the
limitations, there are two robust predictions from these models that
may be confirmed by more complete models or even by
observations: (1) there is a maximum planetary mass, $\sim 2\ME$
beyond which conditions to develop strong and long-lived PMF decline;
(2) the formation of a solid inner core is favored in the case of
$\approx O(1) M_\oplus$ planets.  These two features are of
fundamental importance to our results.  Changes in other outputs of
the thermal evolution model such as the differences between the
convective power for planets with different mass or the role that
other planetary properties will have in the onset of a dynamo, will
not change the conclusions of this work.

Hypothesis (2) is also a matter of concern in present studies of
planetary and stellar dynamos \citep{Christensen10}.  Although the
application of numerical dynamo-based scaling laws to planetary
dynamos has had some success \citep{Olson06}, higher resolution in
future numerical experiments aimed at exploring a wider region of the
parameter space is required to confirm this hypothesis.  Further
advances in the understanding of how non dipolar dynamos behave will
also be required to support the procedure devised in this work to
estimate the PMF intensity in that case.

Hypothesis (3) relies again on inferences from extensive numerical
dynamo experiments (see Fig. \ref{fig:bdip-fdip}).  Since the
numerical parametric studies of \citet{SJ06} it is well known that
dipolarity decreases with the increase of the ratio of inertial to
Coriolis forces.  \citet{Christensen06} identified a critical value
for $\Rol$, this critical behavior has been confirmed by more recent
studies performed by \citet{Aubert09}, \citet{Driscoll09} and
\citet{Christensen10}.  \citet{Driscoll09} found a threshold in $Ra_Q$
separating the dipolar and multipolar regimes (see Fig. 3b in their
paper) consistent with the $\Rol$ critical value found by
\citet{Christensen06} and \citet{Aubert09}.  Less clear are the
properties of the dynamos lying close to the boundary between the
dipolar and multipolar regimes.  The best known planetary dynamo,
i.e. the Earth's dynamo, is just right there.  \citet{Driscoll09}
discussed the relationship between the particular reversal history of
the Earth and the unknown properties of the transitional region
between dipolar and multipolar regimes in parametric space.  In this
case however an open question remains: why is the Earth's dynamo so
close to this boundary? The scaling law for $\Rol$ found in
Eq. (\ref{eq:Roml-MtP}) could shed light into this ``coincidence
problem''.  It is noted that the particular thermal history of our
planet does not affect to a large extent the order of magnitude of
$\Rol$.  As shown in Fig. \ref{fig:Roml-t-Po24} and \ref{fig:p-t},
one or two orders of magnitude variation of $p$ are not enough to
change the order of magnitude of $\Rol$.  This quantity is more
sensitive to the period of rotation and the core radius of the planet.
The period of rotation of the Earth has been of the same order since
the formation of the planet \citep{Denis11} and it is close to that of
Mars.  The core radius is mainly determined by the Fe/Si ratio, a
quantity that is not ``fine-tuned'' for the Earth since Venus and Mars
has a similar value of this ratio.  In conclusion the present value of
the Earth's dynamo local Rossby number is not just coincidentially
close to the boundary region since, as argued here, reasonable
variations in the key properties of the dynamo will place it not too
far from this region.



It should be stressed that after the results presented in this work
and taking into account the fact that probably most of the habitable
SEs lie in tidally locked regions (see e.g. \citealt{Boss06} and
\citealt{Forveille09}) where large periods of rotation could be
common, future efforts to try to understand the emergence and
evolution of PMFs in these types of planets, must consider the kind of
direct and indirect effects that rotation have in the PMF properties
as those considered here.  In other words the assumption of rapidly
rotating planets, i.e. $P\lesssim 2$ days, is no longer valid if we
want to study tidally locked SEs inside the HZ of M-dwarfs.  On the
other hand assuming that tidally locked planets lacks completely of an
intense planetary magnetic field is also an oversimplification.  As
has been shown here there are a range of periods of rotation where
planets could sustain moderate magnetic fields having large periods of
rotation.

Thermal evolution models used in this work assumed planetary
properties very similar to Earth.  \citet{Gaidos10} studied the impact
that several modifications to this reference model will have on their
results.  An interesting case is that of planets with a different
relative core size.  They found that an increase in core size (a
larger Fe to Si ratio) essentially has two effects: (1) an earlier
nucleation of the solid inner core and (2) an increase of the surface
magnetic field intensity.  The latter effect is mostly due to the
smaller attenuation of core field and not to a noticeable modification
of the convective power density.  A different core size changes our
results in two important ways: (1) the predicted maximum dipolar
component of the PMF is increased and (2) the dipolar field lifetime,
especially in the case of slow rotators, is decreased.  Assuming that
a different core size does not affect noticeably the convective power,
the local Rossby number will be slightly modified and hence the
general results regarding, for example, the intervals of rotation
periods for the new categories of rotators, will not be substantially
altered.

\section{Summary and conclusions}
\label{sec:conclusions}

We studied the role of rotation in the evolution of dynamo-generated
magnetic fields in Super Earths.  We computed the evolution of the
local Rossby number and the volumetric magnetic field strength for
core dynamos in SEs.  For this purpose we used the results of two
recently published thermal evolution models and scaling laws fitted
with numerical dynamo experiments.  Assuming that the local Rossby
number could be used as a proxy to dynamo regime, we estimated the
maximum dipolar component of the magnetic field at the CMB, and from
it an upper bound to the dipolar part of the field at the planetary
surface.

We used two properties to characterize the global magnetic properties
of SEs: (1) the average of the surface dipolar component of the field,
$B_{avg}$ and (2) the total time $T_{dip}$ spent by the dynamo in the
dipolar dominated regime (reversing and non reversing).  Intense
magnetic fields with a strong dipolar component
\citep{Stadelmann10}, are best suited to protect planetary
environments from external agents (stellar wind and cosmic rays).
Therefore large values of $B_{avg}$, irrespective of the dynamo
regime, are consistent with planetary habitability.  The long-term
preservation of water and other volatiles in a planetary atmosphere
and the development of life, would require long-lived protective PMFs,
i.e. large values of $T_{dip}$.  Intense and protective magnetic
fields in the early phases of planetary and stellar evolution will be
also suited for the preservation of an atmosphere or its volatiles.
However a planet that achieves to preserve its atmosphere during the
harsh conditions of the early active phases of stellar evolution but
lacks of a protective magnetic field soon after this period will leave
emerging forms of life to an integrated effect of galactic and stellar
cosmic rays induced damages.

We found that the PMF properties depend strongly on planetary mass and
rotation period.  Rapidly rotating SEs $P\approx O(1)$ day, with mass
$\approx O(1) \ME$, have the best potential to develop long-lived and
intense PMFs.  More massive planets develop weaker magnetic fields but
they have dipolar dominated dynamos in a slightly larger range of
rotational periods $P\sim 1-3$ days.  SEs with rotation periods larger
than $3-10$ days (depending on their mass) will spend the majority of
the dynamo lifetime in a multipolar state.

In order to summarize our results we have introduced a rotation-based
classification.  Using the \CTE thermal evolution model, SEs could be
rapid rotators $P\lesssim 1.5-4$ days, slow rotators $4\lesssim
P\lesssim 10-20$ days and otherwise, very slow rotators.  Planets in
the HZ of low mass stars $\Ms<0.6$ that will be tidally
locked in less than 1 Gyr, will fall between the slow and very slow
rotator types.  Unlocked planets could be any of the types described
before, according to their primordial period of rotation and the
effects that could dampen it.

More theoretical and observational efforts should be undertaken to
address the problem of direct or indirect detection of PMF around low
mass planets.  The detection and measurement of such planetary
magnetic fields will help us to constrain thermal evolution and dynamo
models.  The role of magnetic fields in planetary habitability is
another problem that deserves close attention.  Recent works have
tackled this problem in detail \citep{Griebmeier10} but their PMF
models are too simplistic.  Although the effect of rotation rates is
considered in those models and they have focused on tidally locked
planets, their models do not include the effects of thermal evolution
on the PMF properties and their treatment of the dependence of these
properties of the rotation rate is also limited.


\section*{Acknowledgments}

We want to thank D. Valencia for encouraging us to complete this work
and for her useful comments and discussion about our first
approximations of the problem.  We also thank S. Labrosse and
U.R. Christensen for their kind answers to our questions on numerical
dynamo experiments and thermal evolution of the Earth's core.  Special
thanks to E. Gaidos who kindly provided us with details of the results
published in \citet{Gaidos10}.  U.R. Christensen kindly shared with us
the results of tens of numerical dynamo experiments that were
fundamental to improve our results and to confirm our conclusions.  We
are also grateful with Luz Angela Cubides and Luke Webb for the final
revision of the manuscript.  Finally we thank the anonymous referees
who made so many useful comments on the content and about the style of
the manuscript which at the end led to its final form.  PC is
supported by the Vicerrectoria de Docencia of the Universidad de
Antioquia.  This work has been done with the financial support of the
CODI-UdeA under Project IN591CE and the Universidad de Medellin under
the Project CIDI479.

\bibliographystyle{icarus}
\bibliography{bibliography}

\begin{table}[ht]
  \centering
  \begin{tabular}{lll}
    \hline\hline Symbol & Meaning & Notes \\\hline
    \multicolumn{3}{c}{\bf Acronyms} \\\hline
    PMF & Planetary Magnetic Field & Surface magnetic field\\
    CMF & Core Magnetic Field & Core surface magnetic field \\
    \CTE & Core Thermal Evolution & \citet{Gaidos10} \\
    \MTE & Mantle based Thermal Evolution & \citet{Tachinami10} \\
    HZ & habitable zone & \citet{Kasting93} \\
    \multicolumn{3}{c}{\bf Planetary Properties} \\\hline
    $R_p$ & Planetary radius, $R_p=6371 (M/\ME)^{0.265}$ & km, \citet{Valencia06}\\
    $R_c$ & Radius of the core, $R_c=3480 (M/\ME)^{0.243}$ & km, \citet{Valencia06}\\
    $\bar{\rho}_c$ & Average core density, $\bar{\rho}_c=1.1\times 10^4 (M/\ME)^{0.271}$ & kg m$^{-3}$, \citet{Valencia06}\\
    $\Omega$,$T$ & Rotation rate, period of rotation, $T=2\pi/\Omega$ & rad s$^{-1}$, days\\
    $R_{ic},\chi$ & Radius of the solid inner core, $\chi=R_{ic}/R_c$ & km \\
    $D$ & Vertical height of the liquid core, $D=R_c-R_{ic}$ & km \\
    $V$ & Volume of the dynamo region, $V=4/3 \pi (R_c^3-R_{ic}^3)$ & km$^3$ \\
    \multicolumn{3}{c}{\bf Dynamo Properties} \\\hline
    $Q_{conv}$ & Total convective power & $W s^{-1}$ \\
    $p$ & Total convective power density & Adimensional \\
    $Lo$ & Lorentz number, $Lo\sim <E_{mag}>^{1/2}$ & Adim., \citet{Christensen06}\\
    $Ro$ & Rossby number, $Ro\sim <E_{kin}>^{1/2}$ & Adim., \citet{Christensen06}\\
    $Ro_l$ & Local Rossby number, $Ro_l\sim <l_u><E_{kin}>^{1/2}$ & Adim., \citet{Christensen06}\\
    $f_{ohm}$ & Fraction of ohmic dissipation & Adim., \citet{Christensen06} \\
    \multicolumn{3}{c}{\bf Magnetic Field Properties} \\\hline

    $B_{rms}$ & rms amplitude of the magnetic field inside the
    convecting shell & $\mu T$ \\
    $\bar{B}_{dip}$ & Dipolar component intensity of the CMF & $\mu T$
    \\
    $\bar{B}_{s,dip}$ & Dipolar component of the PMF,
    $\bar{B}_{s,dip}=\bar{B}_{dip}(R_c/R_p)^3$ & $\mu T$\\
    $f_{dip}$ & Dipolar fraction of the CMF,
    $f_{dip}=\bar{B}_{dip}/\bar{B}_{CMB}$ & Adim., \citet{Christensen06} \\
    $b_{dip}$ & Ratio between the rms strength of the field \\
     & and the dipolar component at the CMB, 
    $b_{dip}=B_{rms}/\bar{B}_{dip}$ &  Adim., \citet{Christensen09} \\
    $t_{ic}$ & Starting time for the inner core nucleation & Gyrs \\
    $T_{dip}$ & Dipolar lifetime & Gyrs \\
    $t_{sw}$ & Dipolarity switch time & Gyrs \\
    $T_{dyn}$ & Dynamo lifetime & Gyrs \\
    \hline\hline
  \end{tabular}
  \caption{Symbols and quantities used in this work.
  \label{tab:symbols}}
\end{table}

\begin{table}
  \centering
  \begin{tabular}{c|c|cccc}

    \multicolumn{6}{c}{\CTE model ($t_{ic}$,$B_s$,$T_{dyn}$)} \\\hline\hline
    \multirow{2}{*}{$M$ ($\ME$)}  & \multirow{2}{*}{Tectonics} & \multicolumn{4}{c}{$T_s$}\\
    \hfill & \hfill & \mc{288K} & \mc{1500K}\\\hline\hline
    $1\ME$   & PT &  \mc{\bf 2.8, 90, $>$10} & \mc{1.7,140,$>$10} \\
    \hfill   & SL &  \mc{6.5, 0, $>$10} & \mc{5.9, 0, $>$10} \\\hline
    $1.5\ME$ & PT &  \mc{\bf 4.4, 20, $>$10} & \mc{2.7, 130, $>$10} \\
    \hfill   & SL &  \mc{--} & \mc{6.5, 0, $>$10} \\\hline
    $2\ME$   & PT &  \mc{\bf 6.8, 20, $>$10} & \mc{4.2, 90, $>$10} \\\hline
    $2.5\ME$ & PT &  \mc{\bf $>$10, 20, 7} & \mc{$>$10, 30, 6.4} \\\hline
    $3.0\ME$ & PT &  \mc{\bf $>$10, 20, 6.5} & \mc{$>$10, 30, 10}\\\hline
    $4.0\ME$ & PT &  \mc{\bf $>$10, 20, 5.2} & \mc{$>$10, 30, 9} \\
    \hline\hline

    \multicolumn{6}{c}{} \\
    \multicolumn{6}{c}{\MTE model ($t_{ic}$,$B_s$,$T_{dyn}$)} \\\hline\hline
    \mr{$M$ ($\ME$)} & \mr{$V^*$} & \multicolumn{4}{c}{$\Delta T_{CMB}$}\\
    \hfill & \hfill & 1000K & 2000K & 5000K & 10000K \\\hline\hline
    $1.0\ME$ & 3 m$^3$ mol$^{-1}$ & 4, 80, $>$20 & 6.5, 110, $>$20 & 7.5, 130, $>$20 & 7.5, 130, $>$20 \\
    \hfill   & 10 m$^3$ mol$^{-1}$ & {\bf 2.7, 80, 10} & 2.8, 80, 10 & 2.8, 80, 10 & 2.8, 80, 10 \\\hline
    $2.0\ME$ & 3 m$^3$ mol$^{-1}$ & 0, 90, $>$20 & 7, 120, $>$20 & 8, 140, $>$20 & {\bf 8, 140, $>$20} \\
    \hfill   & 10 m$^3$ mol$^{-1}$ & 0, 0, 0.5 & 14, 100, $>$20 & 14, 100, $>$20 & 14, 100, $>$20 \\\hline
    $5.0\ME$ & 3 m$^3$ mol$^{-1}$ & 0, 0, 1 & 7.5, 130, $>$20 & 11, 160, $>$20 & {\bf 11, 160, $>$20} \\
    \hfill   & 10 m$^3$ mol$^{-1}$ & 0, 0, $>$20 & $>$20, 0, $>$20 & $>$20, 150, $>$20 & {$>$20, 150, $>$20}\\
    \hline\hline

  \end{tabular}
  \caption{Summary of results for the evolution of PMF in the \CTE and
    \MTE models \citep{Gaidos10,Tachinami10}.  For every mass and each
    pair of independent planetary properties (tectonics and surface
    temperature, $T_s$ in \CTE model, activation volume, $V*$ and
    temperature contrast at CMB, $\Delta T_{CMB}$ in \MTE model), we
    present the value of three properties of the dynamo and the
    predicted PMF: $t_{ic}$ (Gyrs) the time for the starting of the
    inner core nucleation, $B_s(t_o)$ ($\mu T$) surface magnetic field
    at a reference time taken here as the present age of the Earth,
    4.54 Gyrs and $T_{dyn}$ (Gyrs) the lifetime of the dynamo. All the
    values are approximated and have been used to characterize the
    global conditions to have a protective PMF.  In the \CTE model
    stagnant lid (SL), as opposed to plate tectonics (PT)
    configurations, are not able to produce a dynamo for masses larger
    than 1.5 $\ME$ and are not included in the Table. We have
    highlighted the configurations used in this work to study the role
    of rotation in the PMF evolution.
    \label{tab:thermal-models-results}}
\end{table}

\newpage


\appendix
\section{From the CMF to the PMF intensity}
\label{sec:appendix}

Assuming a low conductivity mantle and neglecting other field sources,
the magnetic field in the region outside the conducting core is
derived from a potential given by the solution of the Laplace equation
with Neumann's boundary conditions at the CMB \citep{Backus96},

\beq{eq:Vexp}
V(r,\theta,\phi)=-\frac{R_c}{\mu_0}\sum_{l=1}^{\infty}\left(\frac{R_{c}}{r}\right)^{l+1}\sum_{m=0}^{l}
a_{l}^{m}Y_{l}^{m}(\theta,\phi) 
\eeq

where $Y_{l}^{m}$ are the spherical harmonics and $a_{l}^{m}$ are the
expansion coefficients.  Using the harmonic expansion, the field
regime (dipolar or multipolar) could be described in terms of the
power spectrum $W_l(r)$ that expands the energy density at radius $r$,

\beq{Wr}
W(r)=\sum_{l=1}^{\infty} W_l(r)\sim\sum_{l=1}^{\infty}\left(\frac{R_c}{r}\right)^{2l+4} w_l 
\eeq

In the case of a dipolar dominated CMF, i.e. a magnetic field where
the averaged dipolar component is a significative fraction of the
total averaged magnetic field strength (see section
\ref{subsec:dynamo-regimes}), the dipolar contribution to the power
spectrum $W_1(R_c)=w_1$ is generally larger than those from higher
order harmonics (see for example the CMF power spectrum of the Earth
and other simulated dynamos in Fig. 4 of \cite{Driscoll09}).
Conversely when the field is multipolar the contribution from $w_1$ is
of the same order or smaller than the higher order harmonics
contributions.  The dependence on $r$ of $W_l(r)$ changes the power
spectrum of the field outside the core.  If the CMF is dipolar
dominated, the surface field will be strongly dipolar.  However, in
the case of a multipolar field the surface field regime will depend on
the CMB power spectrum.  Multipolar fields with a flat spectrum
$w_1\simeq w_l$ for $l>1$ will be dipolar on the planetary surface.  A
strongly multipolar field with a dominated component of order
$l_{max}$ will exhibit a surface flat spectrum provided
$w_{lmax}\simeq w_1 (R_c/R_p)^{2-2l_{max}}$.  Assuming a quadrupole
dominated CMF and $R_c/R_p\approx0.3$ this conditions implies that for
$w_2\lesssim10 w_1$ the surface field will still be weakly dipolar.

The magnetic field outside of the core $\vec{B}(r,\theta,\phi)=-\nabla
V(r,\theta,\phi)$ could also be expanded,

\beq{Br}
\vec{B}(r)=\sum_{l=1}^{\infty}\sum_{m=0}^{l} \vec{B}_{l}^{m}(r) 
\eeq

The contribution of order $l$ is
$\vec{B}_{l}(r)=\sum_{m=0}^{l}\vec{B}_{l}^{m}(r)=\vec{B}_{l}(R_c)(R_c/r)^{l+2}$.
Using this notation the dipolar component of the CMF is written as
$\bar{B}_{dip}=<|\vec{B}_1(R_c)|>$ where the average is computed over
the CMB surface.  On the planetary surface the dipolar component of
the PMF will be then given by,

\beq{Bsdip}
\bar{B}_{s,dip}=\bar{B}_{dip}\left(\frac{R_c}{R_p}\right)^{3} 
\eeq

irrespective of the regime of the surface field.


\end{document}